\documentclass[twocolumn,prx,aps,superscriptaddress]{revtex4-2}
\usepackage{amssymb,amsmath,amsfonts}

\usepackage[latin9]{inputenc}
\usepackage{color}
\usepackage{amstext}
\usepackage{enumitem}
\usepackage{graphicx,bm,palatino}
\usepackage[colorlinks=true,linkcolor=blue,urlcolor=blue,citecolor=blue,pdfusetitle]{hyperref}

\usepackage[sc]{mathpazo} 

\usepackage{hyperref,cleveref}
\usepackage[dvipsnames]{xcolor}
\usepackage[caption=false]{subfig}

\newcommand\ket[1]{\left|#1\right\rangle}
\newcommand\bra[1]{\left\langle #1 \right|}

\usepackage{times}
\usepackage{bbm}

\makeatother

\begin{document}
\title{Efficient excitation-transfer across fully connected networks via local-energy optimization}
\date{\today}

\author{S. Sgroi}
\affiliation{Centre for Quantum Materials and Technologies, School of Mathematics and Physics, Queen's University Belfast, BT7 1NN, United Kingdom}
\author{G. Zicari}
\affiliation{Centre for Quantum Materials and Technologies, School of Mathematics and Physics, Queen's University Belfast, BT7 1NN, United Kingdom}
\author{A. Imparato}
\affiliation{Department of Physics and Astronomy, University of Aarhus,
Ny Munkegade, Building 1520, DK-8000 Aarhus C, Denmark}
\affiliation{Centre for Quantum Materials and Technologies, School of Mathematics and Physics, Queen's University Belfast, BT7 1NN, United Kingdom}
\author{M. Paternostro}
\affiliation{Universit\`a degli Studi di Palermo, Dipartimento di Fisica e Chimica - Emilio Segr\`e, via Archirafi 36, I-90123 Palermo, Italy}
\affiliation{Centre for Quantum Materials and Technologies, School of Mathematics and Physics, Queen's University Belfast, BT7 1NN, United Kingdom}

\begin{abstract}





We study the excitation transfer across a fully connected quantum network whose sites energies can be artificially designed.
Starting from a simplified model of a broadly-studied physical system, we systematically optimize its local energies
to achieve high excitation transfer for various environmental conditions, using an adaptive Gradient Descent technique and Automatic Differentiation.
We show that almost perfect transfer can be achieved with and without local dephasing, provided that the dephasing rates are not too large.
We investigate our solutions in terms of resilience against variations in either the network connection strengths, or size, as well as coherence losses. We highlight the different features of a dephasing-free and dephasing-driven transfer.
Our work gives further insight into the interplay between coherence and dephasing effects in excitation-transfer phenomena across fully connected quantum networks. In turn, this will help designing optimal transfer in  artificial open networks through the simple manipulation of local energies.

\end{abstract}

\maketitle

\section{Introduction}
\label{sec:intro}

Boosted by the unprecedented interest towards quantum information technologies, the study of the properties of complex networks in the quantum domain has received a great deal of attention, due to its broad range of applicability~\cite{Bianconi:2015,Mahler:1998}. Several theoretical studies have indeed shown that modeling complex quantum phenomena in terms of simple quantum systems is not sufficient to capture a plethora of interesting problems belonging to different fields, ranging from -- just to name a few - quantum communication~\cite{Gisin:2007,Chen:2021}, transport phenomena in nanostructures~\cite{Lambert2021,Beenakker:2008},  to quantum biology~\cite{Lambert_review:2013,Huelga_Plenio_quantum_bio:2013}. 

Despite their apparent differences, such phenomena face similar theoretical challenges. On one hand, they require a deeper understanding of the role played by the geometry and topology in the properties of the network itself, as well as its optimal functionality. On the other hand, while exploring quantum dynamical processes in complex networks, it is crucial to assess whether or not genuine quantum features, such as non-classical correlations~\cite{Horodecki_rev:2009}, or genuine quantum processes, such as decoherence~\cite{Zurek_Rev:2003}, may influence the transport properties of a given complex network. The significance of these theoretical studies is essentially twofold: they constitute an attempt to unveil the potential benefits offered by quantum resources, while paving the way towards a better understanding of the way quantum complex networks can be realised in practice.

In this work, we address the problem of identifying a network configuration compatible with optimal transport performances, including the effect of non-trivial interaction between the complex system and an external environment. These instances call for an extensive use of the open quantum systems formalism, the latter being able to effectively describe and interpret the dynamical evolution of a system undergoing irreversible processes such as dissipation and dephasing, which result from the interaction with the environment~\cite{Breuer-Petruccione,Rivas2012,deVega_Alonso_rev:2017}. 

However, attacking this multifaceted issue from the most general standpoint would be a formidable task. We therefore focus on a specific model of open quantum network, whose features have been extensively studied. The latter has been used to effectively describe the phenomenological dynamics of the Fenna-Matthews-Olson (FMO) protein complex~\cite{Engel:2007,Fleming:2009}. This complex plays a pivotal role in light-harvesting process of green-sulphur bacteria: it mediates the highly efficient transfer of excitations from large antenna structures to reaction centres~\cite{Lambert_review:2013,Fleming:2009,LHC_rev:2018}.
The dynamics of such a complex has been modeled and thoroughly studied resorting to the open quantum system paradigm in a series of seminal papers~\cite{Plenio:2008,Caruso:2009,Chin:2013}.
In particular, in this work we refer to Ref.~\cite{Caruso:2009}, where the FMO complex dynamics is described by a network simultaneously undergoing a Hamiltonian dynamics, accounting for the coherent exchange of excitations between the network sites, along with dephasing and dissipative Lindbladian dynamics, leading to loss of coherence and excitations instead. Interestingly, the dynamics exhibits a behaviour, which, to some extent, seems conterintuitive: unlike a classical random walk model, whenever one studies the fully Hamiltonian dynamics (i.e., in absence of dephasing and dissipation), the transport through the network can be inhibited as a consequence of destructive interference between sites~\cite{Caruso:2009}. Such destructive interference can be suppressed by either adding local static disorder -- eventually leading to perfect excitation transfer in the limit of random local energies -- or adding local dephasing noise. The addition of static disorder contradicts the celebrated Anderson localization, according to which random disorder is responsible for inhibition of fully coherent transfer~\cite{Anderson:1958,Anderson_rev:1978}. The effects of local dephasing mechanisms, instead, clearly show that this is a relevant example of environment-assisted transport~\cite{Plenio:2008,Mohseni:2008,Rebentrost:2009}: contrary to expectations, the effect of dephasing is not necessarily detrimental for the performance of transport, which could instead be enhanced in certain conditions~\cite{Plenio:2008,Caruso:2009}. 

Our work systematically explores the interplay between optimal transport and different instances of dephasing noise affecting the system's coherence.
More specifically, we focus on the model introduced in Ref.~\cite{Caruso:2009} to describe the FMO complex, where the latter is represented by a $N=7$-site fully connected network (FCN), i.e, a network where each site is connected to any other one. We find the optimal distribution of local on-site energies resulting in the best population transfer  -- constrained by the interaction strengths as gathered from experiments~\cite{Adolphs:2006} --  under different dephasing conditions.

Focusing on the local energies without changing the interaction strengths greatly simplifies both the numerical optimization and practical implementation of the resulting network. Furthermore, in similar systems, there is evidence that the excitation transport efficiency is more susceptible to changes to the site energies than to the connection strengths~\cite{PhysRevResearch.3.L032001}.
We also assess how robust the transfer performance is against changes in the network configuration. In particular, for fixed $N$, we perform arbitrary changes in the network connectivity, and also change the initial site, where the excitation is initially injected, or the final site where the excitation is extracted from the network (the so-called sink). 
In all these cases we find that the presence of even a moderate local dephasing noise makes the transport properties quite robust against arbitrary change in the network properties.

We then go beyond the prescriptions imposed by
experimental data and perform the optimization step by choosing random (albeit fixed) coupling strengths. This is consistent with the aim of this work that, as we should indeed stress, is not about ascertaining how effective the FCN representation of  Ref.~\cite{Caruso:2009} is at correctly reproducing the FMO phenomenology. Rather, we would like to address the potential to improve transport performances in a given network, whose architecture is well justified by experimental evidence, by exploiting its quantum features. 

The remainder of the paper is organised as follows. In \Cref{sec:Model}, we describe our model of open $N$-site FCN, whose dynamics 
is affected by Markovian dissipation and dephasing. In \Cref{sec:Methods}, we give more details about the methodology used throughout the paper. In particular, we discuss the standard procedure for Markovian master equation vectorization used  both for numerically simulating the system dynamics, and arranging the parameters over which performing the optimization in a suitable way. Using an adaptive gradient-descent technique, we run the numerical simulations whose results and analysis are given in \Cref{sec:Results}. Focusing on the case of a network made of $N=7$, we thoroughly study its perfomances by optimizing the local energies, while the couplings between the network sites are given. In the same Section, we  discuss the resilience of the network against changes in the network configuration.  We finally draw our conclusions in \Cref{sec:conclusions}, where possible future directions are also discussed.

\section{Description of the model}
\label{sec:Model}

\begin{figure}
	\centering\includegraphics[width=\columnwidth]{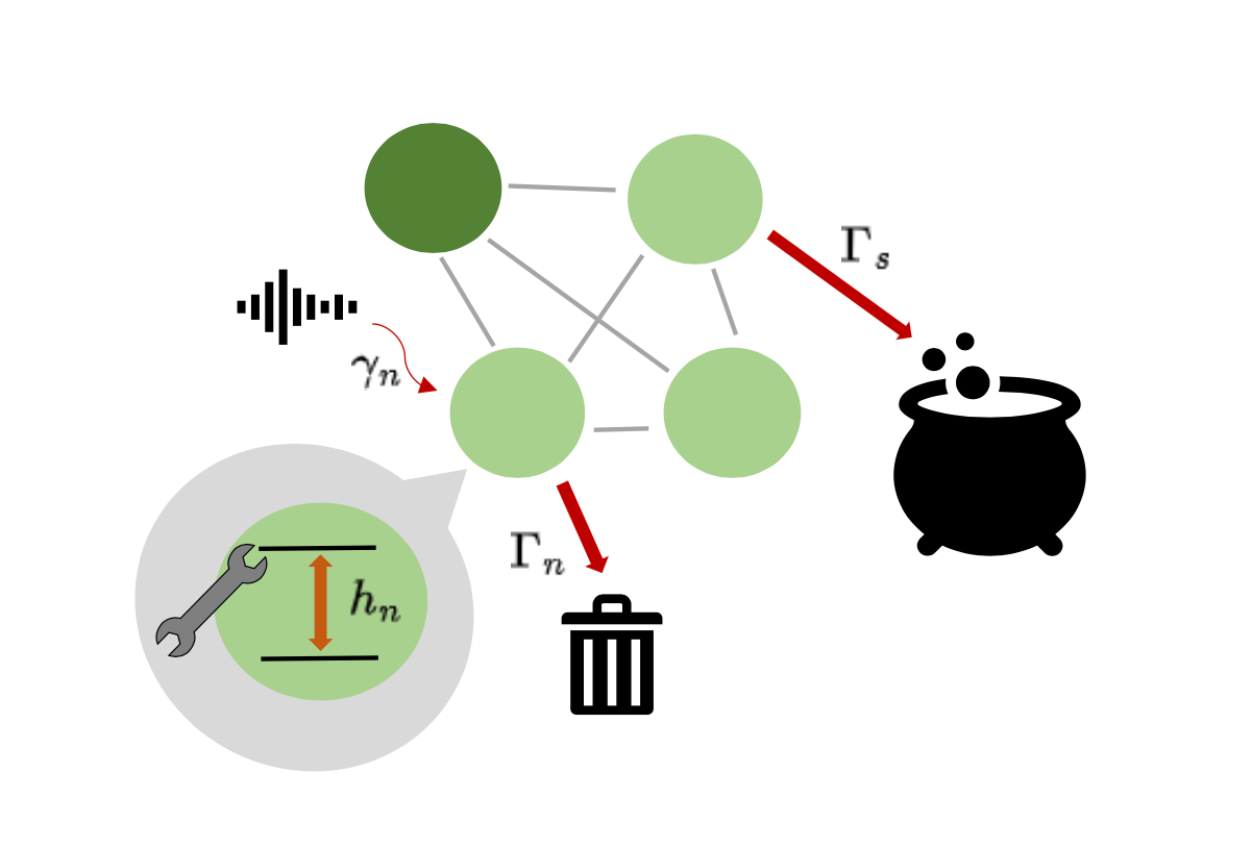}
	\caption{Sketch of the physical situation investigated in this paper. We consider a fully connected network comprising $N$ sites. The excitation is initially injected in one site of the network (darker green), while a different site is attached to a sink into which the excitation is transferred (at a rate $\Gamma_s$). The generic $n$-th site of the network is locally affected by dephasing noise (at a rate $\gamma_n$) and  spontaneous emission, the latter causing the excitation to be irreversibly lost at rate $\Gamma_n$. By optimizing the set of local energies $h_n$, we systematically study the efficiency of excitation-transfer to the sink  under different dephasing conditions.}\label{Figure:scheme}
\end{figure}

Following Ref. \cite{Caruso:2009}, we consider a FCN made of $N$ sites --- Cf. \Cref{Figure:scheme}. We assume that, together with the Hamiltonian dynamics, the system is affected by two different noise mechanism: local pure dephasing, which destroys coherence of any superposition of states, and local spontaneous emission, causing the network to irreversibly transfer  excitations from one site to the environment. We further assume that one excitation at a time can be transferred across the network, i.e., we work in the single-excitation subspace.
This assumption reduces the complexity of the problem, as we scale down the Hilbert space dimension from $2^N$ to $N$, still exhibiting interesting physics. In order to fix the notation, let us introduce the basis $\{\ket{n} \}$ ($n=1,\ldots, N$). In this basis, the unitary dynamics of the FCN is captured by a $N\times N$ Hamiltonian containing the energies associated to each site, as well as the coupling between them. For purposes that will become clear in the following Sections, the system Hamiltonian can be decomposed as
\begin{equation}
\label{eq:full_H}
H=H_D+H_I,
\end{equation}
where $H_D$ is the diagonal part containing all the on-site energies, while $H_I$ contains the coupling between any two sites of the network. The former can be decomposed as
\begin{equation}
\label{eq:HD}
    H_D=\sum_n h_n|n\rangle \langle n|,
\end{equation}
where $h_n$ is the energy associated to the $n$-th site, while the latter is given by
\begin{equation}
    \label{eq:HI}
    H_I = \sum_{m,n} J_{mn} \ket{m} \! \bra{n},
\end{equation}
which, in the language of graph theory, represents the so-called adjacency matrix~\cite{Barabasi_rev:2002}. Note that, as we are dealing with a FCN, $H_I$ is not a sparse matrix, meaning that, in general, we have $J_{mn}\neq0$.

The whole picture is actually completed by introducing two auxiliary sites to the network: one where excitations are irreversibly lost after spontaneous emission, the other where excitations are transferred to, and which mimics the reaction center in photosynthetic complexes such as the FMO complex. This site is named, from now on, the \emph{sink}. Owing to such extra sites, we are actually working with an $(N+2)$-dimensional Hilbert space, therefore we complete our basis by introducing $\ket{0}$ and $\ket{s}$, which identify the aforementioned extra sites, respectively. 

Resorting to this notation, the pure dephasing process is formally described by local Lindblad operators of the form
\begin{equation}
\label{eq:Lindlad_ops_deph}
    L_{\gamma_n}=\sqrt{\gamma_n}|n\rangle \langle n|,
\end{equation}
where $\gamma_n$ is the dephasing rate. Differently, the spontaneous emission processes are modeled through the set of Lindblad operators
\begin{equation}
\label{eq:Lindblad_ops_relax}
    L_{\Gamma_n}=\sqrt{\Gamma_n}|0\rangle \langle n|,
\end{equation}
where $\Gamma_n$ is the rate with which the excitation is lost in the local environment. As we said earlier, we introduce a sink, where the excitation travelling through the network is transferred with a rate $\Gamma_s$. Similarly to \Cref{eq:Lindblad_ops_relax}, this process is physically modeled as a spontaneous emission, whose associated Lindblad operator reads 
\begin{equation}
    L_{\Gamma_s}=\sqrt{\Gamma_s}|s\rangle \langle m|,
    \label{eq:L_Gamma_s}
\end{equation}
$\ket{m}$ being a given site of the network, i.e., $m = {1, \ldots N}$. Note that the latter ensures that population is irreversibly transferred to the sink once the target site $\ket{m}$ is reached.

We assume that our system undergoes a fully Markovian irreverisble dynamics, therefore the corresponding Lindblad master equation reads 
\begin{align}
\label{eq:Lindblad}
    \frac{d\rho}{dt}= & -\frac{i}{\hbar}[H, \rho] \nonumber  \\
    & + \frac{1}{2} \sum_{\substack{\mu= \{\gamma_n\}, \\  \{\Gamma_n\}, \\ \Gamma_s}} \left(2 L_\mu\rho L^{\dagger}_\mu-L^{\dagger}_\mu L_\mu \rho  -\rho L^{\dagger}_\mu L_\mu\right),
\end{align}
where $H$ is suitably defined over the enlarged $(N+2)$-dimensional Hilbert space, while the sums over $\gamma_n$ and $\Gamma_n$ are meant to run over all the possible values of $n= 1, \ldots, N$. Note that the case of a uniform network, i.e., when $h_n, J_{mn}, \gamma_n, \Gamma_n$ are all equal for any value of $n$, can be analytically solved --- Cf. Appendix A in Ref. \cite{Caruso:2009}. In a more general setting, one can include temperature and memory effects by replacing \Cref{eq:Lindblad} by a more general master equation, as done, e.g., in Ref. \cite{Chin:2013} using the numerically exact Time Evolving Density with Orthogonal Polynomial Algorithm (TEDOPA)~\cite{Chin:2010,Prior:2010,Tamascelli:2019}. However, including these effects is beyond the scope of this work, so we restrict to the Markovian master equation in Eq.~(\ref{eq:Lindblad}). Finally, it is worth stressing that we have made the underlying assumption of local environmental mechanisms. This adheres well with a scenario where the nodes of the network are spaced more than any spatial correlation-length of the environment. While this allows to explicitly bypass the possibility of environment-induced effects in the transport of the excitations to the sink, it matches the situation encountered in situations of simulated networks consisting of matter-like information carriers effectively connected by radiation-based quantum buses and addressed by local potentials to tune their respective local energies. Although identify a specific arrangement is not among the goals of our investigation, we will have such an architecture in mind, implicitly, in the remainder of our formal analysis. 

\section{Methods}\label{sec:Methods}

Given the model introduced in \Cref{sec:Model}, our goal is to optimize the local on-site energies $\vec{h} \equiv \{h_1, \ldots h_N \}$ in order to further improve the excitation transfer, under different dephasing conditions.

First, prior to the optimization problem, we need to solve the system dynamics. To this end, one widely used option is to vectorise \Cref{eq:Lindblad} \cite{vectorization:2015}. Note that, by construction, the reduced density operator is represented by a $(N+2) \times (N+2)$ Hermitian matrix $\rho$, which also automatically encodes information about two extra sites introduced in \Cref{sec:Model}. Through vectorisation, the density matrix is readily transformed to a $(N+2)^2$-dimensional vector (whence the name of the technique)
\begin{equation}\label{eq:r}
    \rho\rightarrow\Vec{r}=(\rho_{00},\rho_{01},...,\rho_{0 N+1}, \rho_{10},...,\rho_{N+1 N+1}).
\end{equation}
Analogously, the unitary part can be remapped according to
\begin{equation}\label{eq:TH}
    [H,\rho] \rightarrow \mathcal{L}_U\Vec{r} \equiv (I\otimes H-H^T\otimes I)\Vec{r},
\end{equation}
whereas the dissipative part transform as
\begin{align}
\label{eq:Lindbladian_vec}
    &L_\mu\rho L^{\dagger}_\mu-\frac{1}{2}\{L^{\dagger}_\mu L_\mu, \rho\}\rightarrow\mathcal{L}_{\mu}\Vec{r}= \nonumber \\ & \Bigg[(L^\dagger_\mu)^T\otimes L_\mu-\frac{1}{2}(I\otimes {L}^\dagger_{\mu} {L}_{\mu}+({L}^\dagger_{\mu} {L}_{\mu})^T\otimes I)\Bigg]\Vec{r}.
\end{align}
By applying this set of rules, one obtains a first-order differential equation of the form 
\begin{align}
\label{eq:Lindblad-vec}
    \dot{\vec{r}} = -\frac{i}{\hbar}\mathcal{L} [ \vec{r}  (t) ],
\end{align}
where the full vectorised Lindbladian is given by $\mathcal{L}\equiv \mathcal{L}_U + i \hbar \sum_\mu \mathcal{L}_{\mu}$.
Owing to this representation, the system state at a generic time $t$ is formally obtained by exponentiation, i.e.,
\begin{equation}
\label{eq:Lindblad-vec-solution}
    \Vec{r}(t)=e^{-\frac{i}{\hbar} t \mathcal{L}} \; \Vec{r}(0).
\end{equation}
Therefore the sink population is easily obtained considering the $(N+2)^2$-th component of $\Vec{r}(t)$, i.e., $r_s^t=\Vec{r}(t)\cdot\Vec{s}$, where $\Vec{s}=(0,0,...,0,1)$ is a $(N+2)^2-$dimensional vector.
Notice that, after applying the transformation given by \Cref{eq:TH}, we are still able to separate between an interaction term $\mathcal{L}^I_U$ and a term which depends solely on the local energies $\mathcal{L}^D_U$. The latter can be written as
\begin{equation}
\label{eq:HD_vec}
    \mathcal{L}^D_U=\sum_n h_n \mathcal{H}_n,
\end{equation}
with
\begin{equation}
\label{eq:Hn}
    \mathcal{H}_n=(I\otimes |n\rangle\langle n|-|n\rangle\langle n|\otimes I).
\end{equation}
Let us now fix the total evolution time $T$ and prepare the system with an excitation in the $n$-th site. 
The goal of optimizing the population transfer can be achieved by minimising the cost function
\begin{equation}
    C(\Vec{h})=1-r^T_s(\Vec{h}),
\end{equation}
where $\Vec{h}$ represents the set of parameters over which the optimization is performed.

Using \Cref{eq:r,eq:TH,eq:Lindbladian_vec,eq:Lindblad-vec,eq:Lindblad-vec-solution,eq:HD_vec,eq:Hn}, we are able to obtain the sink population $r^T_s$ and calculate its gradient with respect to $\Vec{h}$. The latter can be efficiently done by using Automatic Differentiation techniques~\cite{tensorflow:2015}. We can hence minimise $C(\Vec{h})$ using gradient-based techniques, eventually finding the optimal on-site energies configuration, i.e., $\Vec{h}_{\text{opt}}$. In this work, we chose to use a Root Mean Square Propagation (RMSprop) algorithm~\cite{https://doi.org/10.48550/arxiv.1609.04747}, an an adaptive learning-rate optimization algorithm developed to tackle limitations of stochastic gradient descent in training deep neural networks. It adjusts learning rates for each parameter and divides the gradients by an exponentially weighted moving average of the squares of the derivatives in the parameters updates. This aids convergence, speed, and stability. While reuqiring careful hyper-parameter tuning, RMSprop is a valuable tool for training neural networks, particularly useful for non-stationary objectives and recurrent neural networks.  Our choice is based solely on the fact that we generally observed a comparatively faster convergence to the solution compared to other similar techniques in our numerical simulations.

\section{Analysis and Results}
\label{sec:Results}

As mentioned above, we start our analysis by considering a specific network made of $N=7$ sites, which, according to the evidence experimentally gathered in~\cite{Adolphs:2006}, reproduces quite accurately the excitation transfer operated by a FMO complex. In order to perform the optimization, we assume that the coupling between the network sites are those  given in Ref.~\cite{Caruso:2009}, which, in turn are based on the experimental results given in Ref.~\cite{Adolphs:2006}.
Therefore, the non-diagonal part of the system Hamiltonian is given by
\begin{equation}\label{eq:interactions}
\begin{split}
&H_I =\\
&
\begin{pmatrix}
0 & -104.1 & 5.1 & -4.3 & 4.7 & -15.1 & -7.8\\
-104.1 & 0 & 32.6 & 7.1 & 5.4 & 8.3 & 0.8\\
5.1 & 32.6 & 0 & -46.8 & 1.0 & -8.1 & 5.1\\
-4.3 & 7.1 & -46.8 & 0 & -70.7 & -14.7 & -61.5\\
4.7 & 5.4 & 1.0 & -70.7 & 0 & 89.7 & -2.5\\
-15.1 & 8.3 & -8.1 & -14.7 & 89.7 & 0 & 32.7\\
-7.8 & 0.8 & 5.1 & -61.5 & -2.5 & 32.7 & 0
\end{pmatrix}.
\end{split}
\end{equation}
Here and in the following energy values are expressed in units  of $1.2414 \cdot 10^{-4} \ \text{eV}$, while times are in $\text{ps}$, as in Refs.~\cite{Caruso:2009,Adolphs:2006}.

A few comments about $H_I$ are now in order. \Cref{eq:interactions} is representative of a network with a high level of connectivity -- all the entries are non-zero, as we would expect for a FCN where every site is coupled to any other site -- where evidently coupling strenghts are site-dependent. We further assume that the sink $\ket{s}$ is attached to the third site, represented by $\ket{3}$. This assumption is actually physically motivated in FMO complexes: experimental evidence suggests that the third site is the one coupled with the reaction centre~\cite{Adolphs:2006}. For the sake of completeness, we mention here that more recent experiments revealed the existence of a $8$-th site in the FMO complex~\cite{Tronrud:2009,amBusch:2011}; however, for our purposes, we mainly consider the case of $N=7$ sites, as the largest system, with some results concerning networks with reduced size discussed in Section~\ref{resilience}.

Although our first aim is to optimize the network transport properties over the on-site energies, we will benchmark our numerical findings against those contained in  Ref.~\cite{Caruso:2009}, where the on-site energies are given by,  $\Vec{h}_{\text{ref}}=(215, 220, 0, 125, 450, 330, 280)$, the decaying rates associated to spontaneous emission are $\Gamma_n=\Gamma=5\cdot 10^{-4}$, $\Gamma_s=6.283$, the optimal local dephasing rates are $\Vec{\gamma}_{\text{ref}}=(0.157, 9.432, 7.797, 9.432, 7.797, 0.922, 9.433)$, while the total evolution time is $T=5$ unless differently stated. 

\begin{figure}
	\centering\includegraphics[width=\columnwidth]{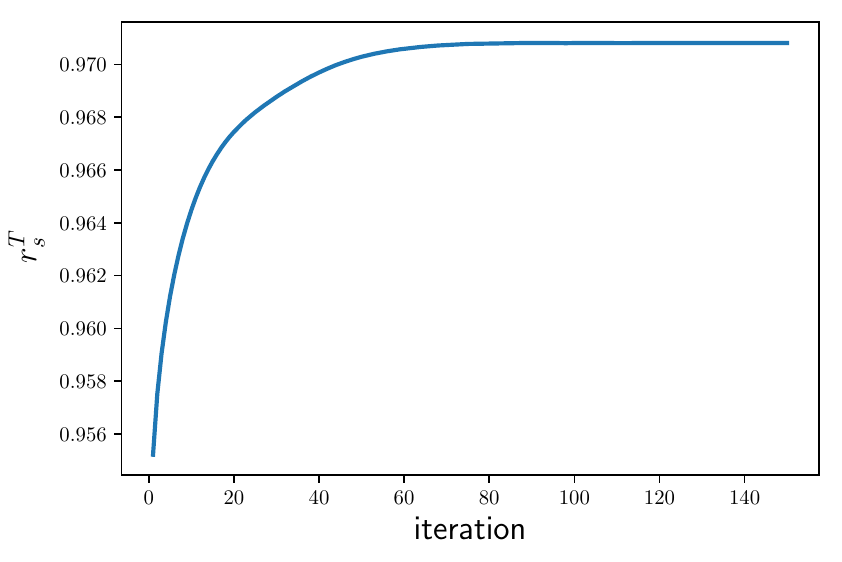}
	\caption{Example of learning curve using the approach presented in \Cref{sec:Methods}.  The final sink population $r_s^T$ is optimized with respect to  $\Vec{h}$, and we plot it as a function of the number of iterations of RMSprop, with $T=5$. The numerical values of the $H_I$ entries are those given by \Cref{eq:interactions}, the decaying rates are given by $\Gamma_n=\Gamma=5\cdot 10^{-4}$, $\Gamma_s=6.283$, while we assume that $\gamma_n=0$ for any value of $n$, meaning that the network is not subject to any dephasing. We assume that the excitation is initially injected in the first site of the network, i.e., $\rho(0)=|1\rangle\langle1|$.}\label{Figure:learning}
\end{figure}

Using the methodology introduced in \Cref{sec:Methods}, we can, for instance, plot a typical learning curve of $\Vec{h}$. In \Cref{Figure:learning}, we show the final sink population $r_s^T$ as a function of the number of the iterations of the optimization algorithm, given a specific network and environment configuration.

\subsection{Optimal solutions}\label{subsec:Optimal solutions}

In this Section, we systematically study the network performance under different dephasing conditions, by looking at the population transferred to the sink over a total evolution time $T$. To this end, we assume that the coupling between the sites is given by $H_I$ in \Cref{eq:interactions}, with the numerical values of the decaying rates as given above. We then study the effectiveness of the optimization of local energies for different dephasing conditions. We indeed start with considering three relevant cases: in the first case, we consider the local dephasing rates $\Vec{\gamma_{\text{ref}}}$ obtained through the optimization performed in Ref. \cite{Caruso:2009}; in the second case, the network sites are not subject to any dephasing, i.e., $\gamma_n = 0$, for any value of $n$; the third case, where the dephasing is uniform across all sites, i.e., $\gamma_n=\gamma = 1$. For easing the notation, we will denote the array of local depashing rates $\Vec{\gamma}$ with $\Vec{\gamma_{\text{ref}}}, \Vec{0}, \Vec{1}$ in the three aforementioned cases, respectively.

For sake of definiteness, we assume that the excitation is initially injected in the first site of the network, i.e., the initial condition for \Cref{eq:Lindblad} reads $\rho(0)=|1\rangle\langle1|$; furthermore we assume that the third site, i.e., $\ket{3}$, is connected to the sink. It is worth mentioning that our results are not qualitatively affected if we change either the state connected to the sink or the initial excited site.

In order to solve the optimization problem, we first initialise the local energies $\Vec{h}_0$ setting them all equal to zero, i.e., $\Vec{h} \equiv \Vec{h}_0 = (0,\ldots, 0)$. By so doing, we obtain the final population transferred to the sink $r^{T}_s(\Vec{h}_0)$. We then perform the optimization over the local energies in the way described in \Cref{sec:Methods} for the three different dephasing conditions, and obtain $r^{T}_s(\Vec{h}^{\Vec{\gamma}}_{\text{opt}})$, with $\Vec{\gamma}=\Vec{\gamma}_{\text{ref}}, \Vec{0}, \Vec{1}$. 

The improvement achieved by optimizing over $\Vec{h}$ can be deduced from the data shown in Table \ref{tab:0vsOpt}, whereas the sink population dynamics is shown in \Cref{Figure:pop_dynamics}. The corresponding optimal Hamiltonians can be found in Appendix \ref{appendix0}. We observe marginal improvement of the population transfer when we take $\Vec{\gamma } = \Vec{\gamma}_{\text{ref}}$, a slight improvement for uniform dephasing rates ( $\Vec{\gamma} = \Vec{1}$) and a larger improvement in absence dephasing, ( $\Vec{\gamma} = \Vec{0}$). In all cases, we are able to achieve high population transfer.

\begin{table}
\begin{tabular}{|c | c | c |} 
 \hline
 $\Vec{\gamma}$ & $r^T_s(\Vec{h}_0)$ & $r^T_s(\Vec{h}^{\vec{\gamma}}_{\text{opt}})$\\ [0.5ex] 
 \hline\hline
 $\Vec{\gamma}_{\text{ref}}$ & $0.955$ & $0.971$\\ 
 \hline
 $\Vec{1}$ & $0.922$ & $0.981$ \\
 \hline
 $\Vec{0}$ & $0.639$ & $0.989$\\ [0.5ex] 
 \hline
\end{tabular}\caption{Final sink population $r_s^{T}$ ($T=5$) for different dephasing consitions, i.e., $\vec{\gamma}=\vec{\gamma}_{\text{ref}}, \vec{1}, \vec{0}$. In the right-hand column we consider the case where all the local energies $\vec{h}$ are set to zero, i.e., $\vec{h}_0 = (0,\ldots, 0)$, while in the left-hand column, we show the final sink population for the optimal local energies $\Vec{h}^{\vec{\gamma}}_{\text{opt}}$ as obtained with the optimization method discussed in \Cref{sec:Methods}.}\label{tab:0vsOpt}
\end{table}

\begin{figure}[t!]
	\centering\includegraphics[width=\columnwidth]{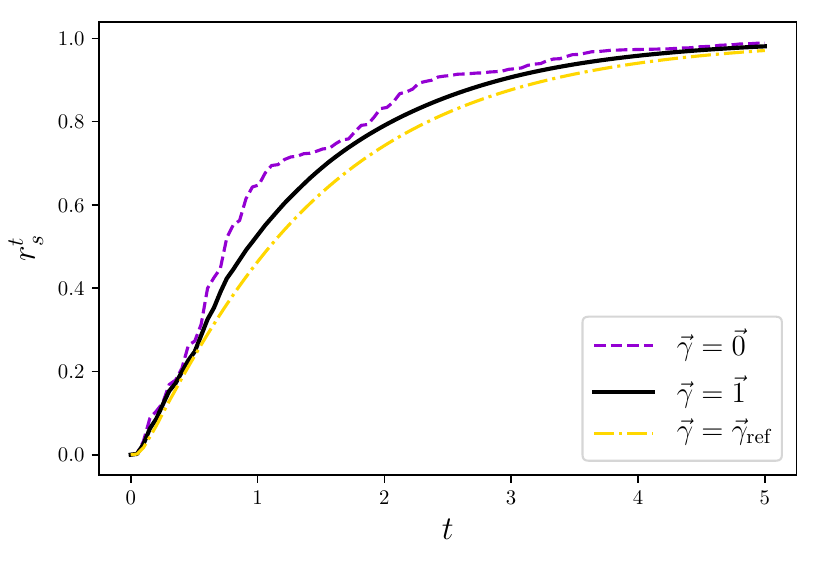}
	\caption{Plot of the sink population $r_s^t(\vec{h}^{\vec{\gamma}}_{\text{ opt}})$ as a function of $t$, where $\vec{h}^{\vec{\gamma}}_{\text{ opt}}$ are the optimal on-site energies obtained through optimization, with the three sets of dephasing rates $\Vec{\gamma}=\Vec{\gamma}_{\text{ref}}, \Vec{0}, \Vec{1}$. The remaining parameters are the same as in \Cref{Figure:learning}. Notice that we are able to achieve high population transfer both with and without dephasing noise.
	}\label{Figure:pop_dynamics}
\end{figure}

\begin{figure}
	\centering\includegraphics[width=\columnwidth]{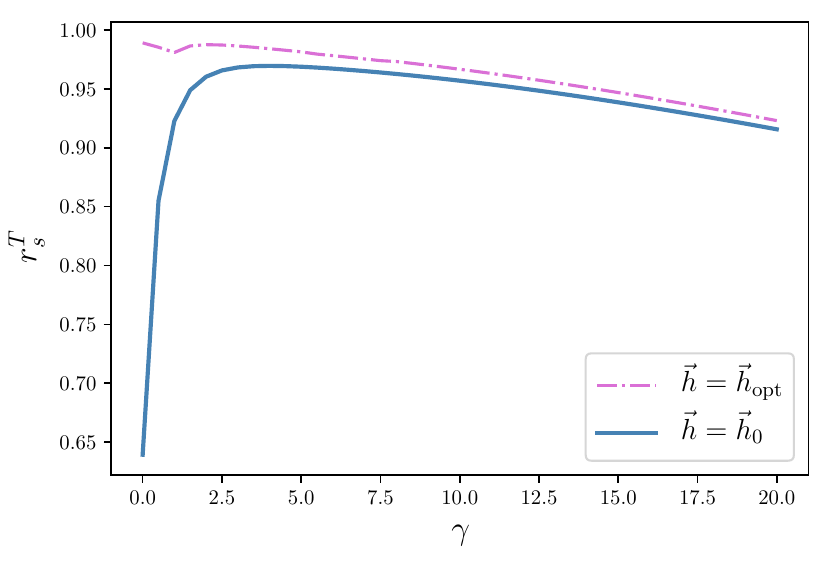}
	\caption{Final sink population $r_s^T$ ($T=5$) for uniform dephasing rate across all network sites $\gamma_n=\gamma$. We compare the case where on-site energies $\vec h$ are the result of the optimization (dash-dotted line) with the case in which we assume them to be all null (solid line). On the one hand the plot shows that the optimization procedure only leads to minor improvements in the population transfer for moderate to large values of $\gamma$ compared to the case where all the local energies are set to zero. On the other hand, the optimization procedure results in a significant improvement of the excitation transfer performance for small values of $\gamma$.   }
	\label{Figure:SinkPop_vs_deph}
\end{figure}

We then explore larger uniform dephasing rates by considering $\gamma_n = \gamma$, where $\gamma$ varies in the range $[0,20]$. As shown in \Cref{Figure:SinkPop_vs_deph}, the optimization allows us to effectively transfer population in a large interval of the chosen range; noticeably, the smaller is the dephasing rate, the larger is the improvement compared to the non-optimized scenario. Moreover, numerical investigations show that, even setting all the local energies to zero, the system achieves high population transfer while we increase the value of the dephasing rates. One can also observe that there is an intermediate range of $\gamma$ where the population transfer is high even without optimization. In this range, the optimization of local energies is superfluous to observe high transfer; the process is mostly guided by dephasing, as in the case where $\Vec{\gamma} = \Vec{\gamma}_{\text{ref}}$.
However, when the dephasing rate becomes too large, it turns out to be detrimental to the transfer; we indeed observe a decrease in the final sink population, both in the optimized and in the non-optimized case. This occurrence can be ultimately justified in terms of the quantum Zeno effect~\cite{Misra:1977,Peres:1980,Facchi:2002}: extreme dephasing conditions tend to freeze the system dynamics~\cite{Rebentrost:2009}.

\begin{figure}
	\centering\includegraphics[width=\columnwidth]{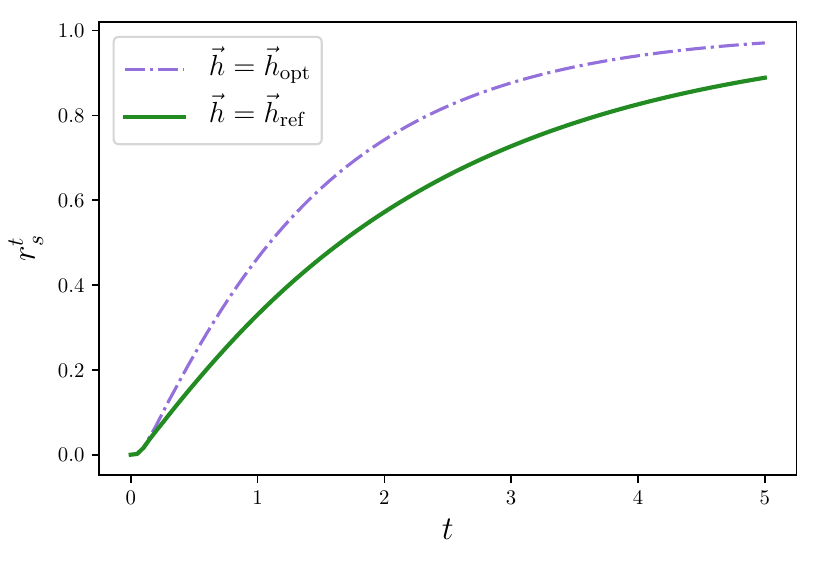}
	\caption{Dynamics of the sink population $r_s^t(\vec{h})$ under the dephasing condition $\Vec{\gamma}=\Vec{\gamma}_{\text{ref}}$. The solid curve corresponds to $\vec{h} = \Vec{h}_{\text{ref}}$, while the dash-dotted curve corresponds to the optimized site energies $\vec{h} = \Vec{h}_{\text{opt}}$. This plot shows that the optimization method discussed in \Cref{sec:Methods} is effective at improving the population transfer.}\label{Figure:comparison_ref}
\end{figure}

To conclude this part of the study we compare our optimal population transfer with the population transfer achieved when taking $\Vec{h} = \Vec{h}_{\text{ref}}$, as given at the beginning of this section. Figure \ref{Figure:comparison_ref} provides evidence of the effectiveness of the optimization: the optimal set of on-site energies $\Vec{h}_{\text{opt}}$ outperforms $\Vec{h}_{\text{ref}}$ for any $t>0$.

\subsection{Resilience against different configurations}\label{resilience}
We now want to test some properties of the optimal on-site energies $\Vec{h}_{\text{opt}}$ for different $\Vec{\gamma}$ discussed in Section \ref{subsec:Optimal solutions}. We start by looking at the resilience of the transfer against variations either in the initial or end sites, or in the coupling between the sites. We consider the optimal solutions for a total evolution time $T=5$, with $H_I$ being given by \Cref{eq:interactions}.
In our analysis the initial and the target sites are always different.
The corresponding results can be found in Table \ref{tab:different_starts_ends}, where in the left-hand column we show the smallest population transferred while varying the site where the excitation is initially injected, while in the right-hand column we show the smallest population transferred when we vary the site connected to the sink. In both cases, the transfer is more resilient when it is mostly guided by dephasing. Indeed, when $\Vec{\gamma} = \Vec{1}$ or $\Vec{\gamma}=\Vec{\gamma}_{\text{ref}}$ we still observe high population transfer, whereas those results are in stark contrast with the zero-dephasing case, when the population transferred to the sink can drop almost to zero.

Next, we look at the effect of allowing population transfer to the sink from a second node of the network. Under this hypothesis, we observe a minimum population of $r_s^{T}\approx0.998$ transferred to the sink, when the latter is connected to both the $3$-rd and $7$-th sites, i.e., $m=3$ and $m=7$ in~\Cref{eq:L_Gamma_s}. Furthermore, it is worth mentioning that we observe no significant differences between the different dephasing conditions.

\begin{table}
\begin{tabular}{|c | c | c |} 
 \hline
 $\Vec{\gamma}$ & $\text{min}(r^T_s)|\rho(0)$ & $\text{min}(r^T_s)|L_{\Gamma_s}$ \\ [0.5ex] 
 \hline\hline
 $\Vec{\gamma}_{\text{ref}}$ & $0.971$ & $0.961$ \\ 
 \hline
 $\Vec{1}$ & $0.981$ & $0.976$ \\
 \hline
 $\Vec{0}$ & $0.018$ & $0.031$\\ [0.5ex] 
 \hline
\end{tabular}
\caption{Smallest population transferred $\min(r_s^{T})$ at $T=5$ for different sets of dephasing noise $\vec{\gamma} = \vec{\gamma}_{\text{opt}}, \vec{1}, \vec{0}$. In the central column, we show those instances in which we vary the initial state $\rho(0)$, i.e., the site where the excitation is initially injected, whereas on the rightmost column we change the site $m$ connected to the sink through the operator $L_{\Gamma_s}$ introduced in \Cref{eq:L_Gamma_s}. }\label{tab:different_starts_ends}
\end{table}

We then use the same local energies while considering different coupling between sites. To do so, we randomly extract the entries of the matrix $H_I$ from a uniform distribution in the range $[-200,200]$. As before, we choose $\ket{1}$ as the initial excited site, while $\ket{3}$ is the target state. Results are shown in Table \ref{tab:different_connections}, where we report the smallest and the largest population transferred to the sink.
We can see that the most resilient transfer is achieved for $\Vec{\gamma} = \Vec{\gamma}_{\text{ref}}$, while, again, the lowest transfer is observed in absence of dephasing noise. It is worth noticing that in all cases we have evidence of a configuration yielding an almost perfect population transfer. The corresponding Hamiltonians can be found in Appendix \ref{appendix0}.

\begin{table}[bp]
\begin{tabular}{|c | c | c |} 
 \hline
 $\Vec{\gamma}$ & $\text{min}(r^T_s)$ & $\text{max}(r^T_s)$ \\ [0.5ex] 
 \hline\hline
 $\Vec{\gamma}_{\text{ref}}$ & $0.833$ & $0.999$ \\ 
 \hline
 $\Vec{1}$ & $0.630$ & $0.999$ \\
 \hline
 $\Vec{0}$ & $0.036$ & $0.999$\\ [0.5ex] 
 \hline
\end{tabular}
\caption{Smallest and largest population transferred $r_s^{T}$ at $T=5$ as obtained with adjacency matrix $H_I$ whose elements are randomly extracted from a uniform distribution defined over the interval $[-200,200]$. We consider $10^4$ realisations of $H_I$, showing that there is at least one matrix $H_I$ leading to almost perfect transfer [cf. right-most column].
}\label{tab:different_connections}
\end{table}

To complete, we study the population transferred to the sink when the network size is reduced. Starting from a FCN of $N=7$ sites, where the adjacency matrix $H_I$ is given by \Cref{eq:interactions}, we progressively scale down the system size, removing one by one the nodes of the network. To this end, we first discard one node of the network (except for the input and the output nodes,  $1$ and $3$, respectively) and update the adjacency matrix by removing the corresponding row and column, then
we optimize over the local energies $\Vec{h}$.  Among all the possible configurations with 6 nodes, we select the one corresponding to the smallest population transferred to the sink after performing the optimization. 
The rationale behind such choice is that, by looking at the worst case scenario, we test the effectiveness of optimizing only the local energies of a smaller network to achieve high excitation transfer.

\begin{figure}

	\centering\includegraphics[width=\columnwidth]{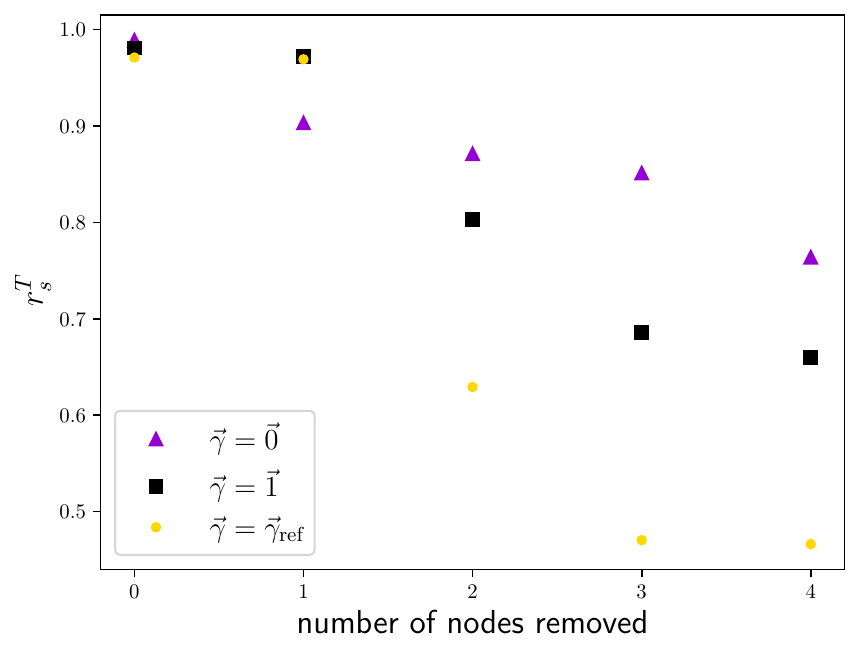}
	\caption{
    Optimized population transferred to the sink $r_s^T$ at $T=5$ as a function of the number of nodes removed from the original network for different dephasing conditions, i.e., $\vec{\gamma}=\vec{\gamma}_{\text{ref}}, \vec{1}, \vec{0}$. See details in \Cref{resilience}.}
	\label{lessnodes}
\end{figure}

We iterate the node-removal followed by optimization procedure until we reach the non-trivial case where we are left with only $3$ nodes.
Results are shown in \Cref{lessnodes}, where it can be seen that this operation has a significant, detrimental impact on the population transfer, showing that carefully selecting the local energies for a given network configuration may not be sufficient to achieve the desired transfer for smaller networks. Furthermore, we can see that, in contrast to changes in the coupling strengths for the seven-site network, reduction of the number of nodes seems to be more detrimental in presence of dephasing noise.

\subsection{Coherence preservation properties}
\label{subsec:coherence}
Results from Section \ref{resilience} have shown that, when the transport is dephasing-assisted, the population transfer is more resilient to changes of the network configuration. On the other hand, we also expect that such process would tend to destroy coherence in the system. This is not necessarily true in absence of dephasing noise, provided that the transfer is fast enough (so that coherence is not destroyed due to excitation losses). 

To study how coherence is preserved or lost from the system during the population transfer, we add a new site $|8\rangle$ to the network, uncoupled from all the other sites. We then prepare the system in the superposition $\frac{1}{\sqrt{2}}(|1\rangle+|8\rangle)$, and we study the time evolution of coherences while the population transfer from site $|1\rangle$ to the target is taking place. 

The irreversible transfer from site $\ket{3}$ to the sink will inevitably lead to coherence loss from the system. However, we would like to separate these artificial losses from the effect of dephasing and spontaneous emission induced by the interaction with the environment. To do so, we instead connect the site $\ket{3}$ to a long spin chain via an interaction Hamiltonian $J_{3 s_0}(|3\rangle\langle s_0|+|s_0\rangle\langle 3|)$, where $\ket{s_0}$ is the first site of the chain. Moreover, the chain is described by the following nearest- neighbour interaction Hamiltonian
\begin{equation}
    H_C=J\sum_{j=0}^{N_C} (|s_j\rangle\langle s_{j+1}|+|s_{j+1}\rangle\langle s_j|),
\end{equation}
where we assume uniform coupling $J$ across the chain, and $N_C$ is the number of sites of the spin chain.

If the chain is long enough and the evolution time considered is not too long, we do not expect revivals to occur, meaning that most of the population transferred to the chain will not go back to the network. This enables us to picture the whole chain as an effective sink. However, in contrast to the previous scenario, the interaction between the network and the chain is affecting the unitary part of the dynamics, therefore it does not induce any additional decoherence. 

\begin{figure}
	\centering\includegraphics[width=\columnwidth]{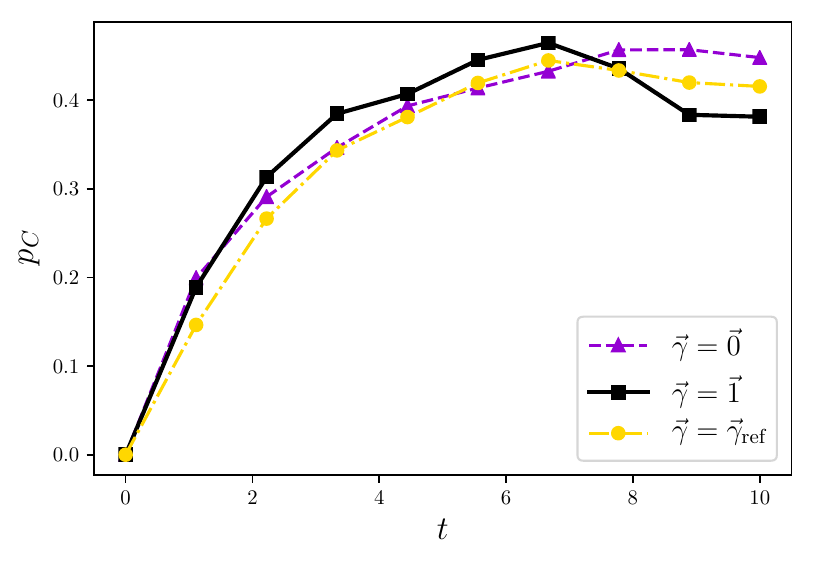}
	\caption{Population $p_{C}$ transferred from the network to the spin chain as a function of time. We consider the optimal site energies $\vec{h}_{\text{opt}}$ (obtained for an evolution time $T=5$), under different choices of the dephasing rates $\vec{\gamma}=\vec{\gamma}_{\text{opt}},\vec{0},\vec{1}$. For the numerical simulations, we chose $N_C=80$ spins in the chain, $J_{3 s_0}/\hbar=\Gamma_s$, and $J=2J_{3 s_0}$, while all the remaining values of the physical parameters are given at the beginning of \Cref{sec:Results}.}\label{Figure:coherencesPopulation}
\end{figure}

We hence study the coherence dynamics in this new scenario for different dephasing rates $\vec{\gamma}$ and the associated optimal on-site energies presented in \Cref{subsec:Optimal solutions}. The total population $p_C$ transferred to the chain as a function of time can be found in \Cref{Figure:coherencesPopulation}. 

\begin{figure}[t!]
	a)\centering\includegraphics[width=\columnwidth]{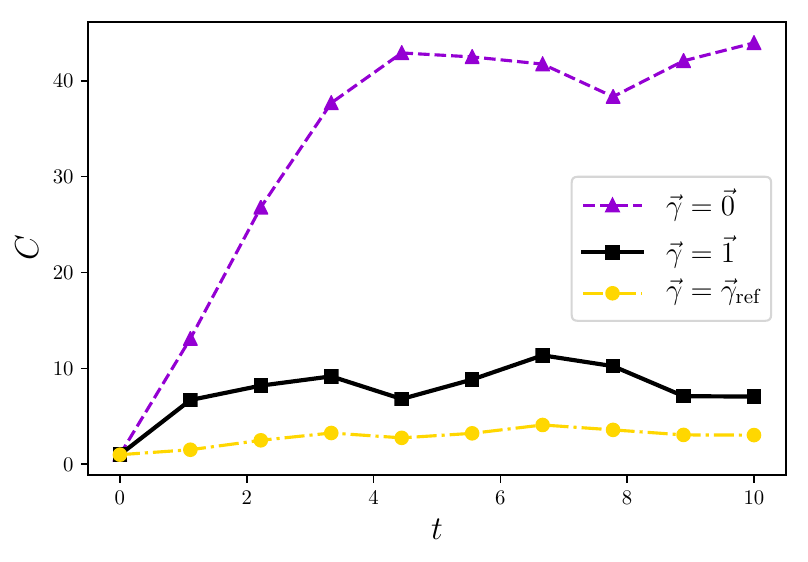}
	b)\includegraphics[width=\columnwidth]{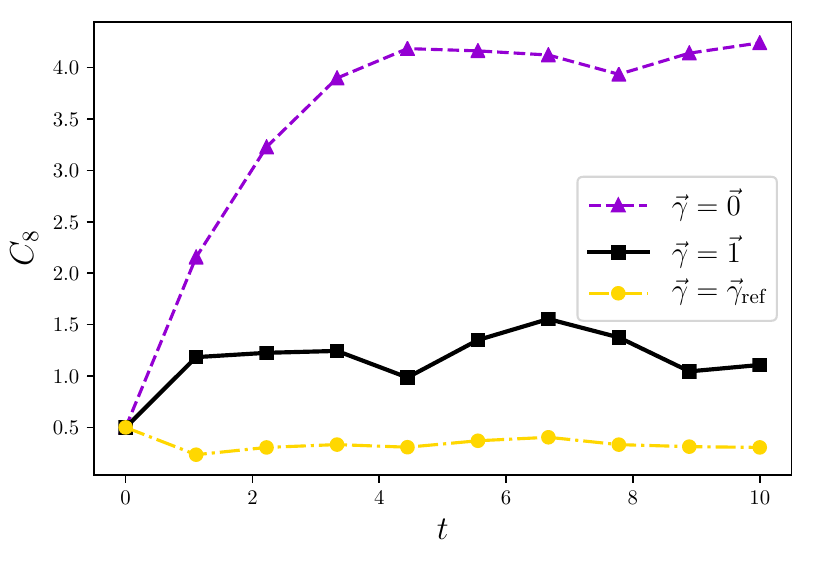}
	\caption{Time evolution of coherence. In Panel (a), we study the system coherence $C$ as quantified by \Cref{Equation:coherence}, while in Panel (b), we look at the reduced coherence $C_8$ associated to the $8$-th site, computed using \Cref{Equation:reduced_coh}.  We compare the curves obtained for different dephasing rates $\vec{\gamma} = \vec{\gamma_{\text{ref}}}, \vec{0}, \vec{1}$, while resorting to the corresponding optimal on-site energies. The parameters used for the numerical simulation are the same as in \Cref{Figure:coherencesPopulation}.}\label{Figure:coherencesCoherences}
\end{figure}

In order to study the time evolution of coherence we employ the standard quantifier given by the $l_1$-norm~\cite{Baumgratz:2014}. In \Cref{Figure:coherencesCoherences}, we show the dynamics of the total coherence of the system computed as
\begin{equation}\label{Equation:coherence}
    C=\sum_{i\neq j} |\rho_{ij}|,
\end{equation}
as well as the coherence associated to the $8$-th site only
\begin{equation}\label{Equation:reduced_coh}
    C_8=\sum_{j} |\rho_{8j}| - \rho_{88},
\end{equation}
where $j=1,..., N, s_0,..., s_{N_C}$. In our simulations, we considered a chain of $N_C=80$ spins, $J_{3 s_0}/\hbar=\Gamma_s$, and $J=2J_{3 s_0}$.

After a time $T=10$, we observe a significant increase in $C$ in absence of dephasing and a small increase when we add dephasing noise. An increase in $C_8$ can also be observed for $\Vec{\gamma}=\Vec{0}$ and $\Vec{\gamma}=\vec{1}$, while $C_8(T)<C_8(0)$ for $\vec{\gamma} = \vec{\gamma}_{\text{ref}}$.
We eventually looked at the coherence per number of sites/spins involved $c=C/N_{\text{tot}}$. For the initial state $c=\frac{1}{2}$, while at the end of the transfer $N_{\text{tot}}=N+N_C$. We obtained $c\approx0.035$ for $\vec{\gamma}_{\text{ref}}$, $c\approx0.080$ for $\vec{\gamma}=\vec{1}$, $c\approx0.499$ for $\vec{\gamma}=\vec{0}$.

These results are in agreement with the expectation that in absence of dephasing, coherence is mostly preserved, while losses can occur when the population transfer is driven by dephasing.

\section{Conclusions and Outlook}
\label{sec:conclusions}

In this work, we optimized the on-site energies of a fully connected quantum network to improve population transfer for different enviromental conditions.
Specifically, we considered a simple model of a FMO complex in the single-excitation subspace, subjected to spontaneous emission and local dephasing. Resorting to a gradient-based technique, we found the optimal site energies for different dephasing rates.
We studied the properties of our solutions in terms of resilience against changes in the network initial preparation, couplings, and size, providing a discussion about coherence preservation during the transfer.
We show that high population transfer in a FMO-like network can be achieved by merely optimizing the sites energies for a large range of different dephasing rates.
However, the optimal solutions for dephasing-driven and zero-dephasing transport are significantly different in terms of resilience to network configuration and coherence preservation.
While in absence of dephasing  we find both a high transport performance, and a high degree of coherence preservation,  
the optimal solutions in the presence of dephasing are shown to be more resilient to changes in the network initial state and couplings between sites, while, as expected, they exhibit a higher loss of coherence in the quantum network state. 
However, in contrast to the transfer resilience against changes in the sites interaction, reduction in the network size seems to be more detrimental in presence of dephasing.

Our work contributes to further understanding of the transport properties of fully connected quantum networks by isolating the effect of local energies, dephasing conditions, and network size for a given set of couplings between the sites.
Furthermore, our results show that a viable and fruitful approach to design efficient synthetic devices is to apply adaptive learning approaches to enhanced existing natural devices, such as the photosynthetic complex that we have considered in this paper.
In this respect, further progress can be made by either studying larger networks, or going beyond the one-excitation subspace, or studying different models of interactions. One might also consider more complex environments, e.g. including non-Markovian effects, as those are displayed by a variety of non-artificial physical systems, and assess whether that could be beneficial for further improving the properties of the transfer.

\acknowledgements

AI gratefully acknowledges the financial support of The
Faculty of Science and Technology at Aarhus University
through a Sabbatical scholarship and the hospitality of the
Quantum Technology group, the Centre for Quantum Materials and Technologies, and the School of
Mathematics and Physics, during his stay at Queen's University Belfast. 

\section{List of abbreviations}
The following abbreviations and acronyms have been used throughout the manuscript:
\begin{itemize}
\item FMO: Fenna-Matthews-Olson 
\item FCN: Fully connected network 
\item TEDOPA: Time Evolving Density with Orthogonal Polynomial Algorithm
\item RMSprop:  Root Mean Square Propagation
\item eV: electron-Volt
\end{itemize}

\section{Declarations}
\subsection*{Ethics approval and consent to participate}
Not applicable.
\subsection*{Consent for publication}
Not applicable.
\subsection*{Competing interests}
The authors declare no competing interests.
\subsection*{ Data availability}
The datasets used and/or analyzed during the current study are available from this \href{https://github.com/SofiaSgroi/Efficient-excitation-transfer-across-fully-connected-networks-via-local-energy-optimization}{link} and from the authors. 
\subsection*{Funding}
We acknowledge support by the European Union's Horizon 2020 FET-Open project  TEQ (766900), the Horizon Europe EIC-Pathfinder project QuCoM (101046973), 
the Leverhulme Trust Research Project Grant UltraQuTe (grant RGP-2018-266), the Royal Society Wolfson Fellowship (RSWF/R3/183013), the UK EPSRC (EP/T028424/1), and the Department for the Economy Northern Ireland under the US-Ireland R\&D Partnership Programme. 

\bibliographystyle{apsrev4-2.bst}
\bibliography{biblio.bib}

\begin{thebibliography}{38}%
\makeatletter
\providecommand \@ifxundefined [1]{%
 \@ifx{#1\undefined}
}%
\providecommand \@ifnum [1]{%
 \ifnum #1\expandafter \@firstoftwo
 \else \expandafter \@secondoftwo
 \fi
}%
\providecommand \@ifx [1]{%
 \ifx #1\expandafter \@firstoftwo
 \else \expandafter \@secondoftwo
 \fi
}%
\providecommand \natexlab [1]{#1}%
\providecommand \enquote  [1]{``#1''}%
\providecommand \bibnamefont  [1]{#1}%
\providecommand \bibfnamefont [1]{#1}%
\providecommand \citenamefont [1]{#1}%
\providecommand \href@noop [0]{\@secondoftwo}%
\providecommand \href [0]{\begingroup \@sanitize@url \@href}%
\providecommand \@href[1]{\@@startlink{#1}\@@href}%
\providecommand \@@href[1]{\endgroup#1\@@endlink}%
\providecommand \@sanitize@url [0]{\catcode `\\12\catcode `\$12\catcode
  `\&12\catcode `\#12\catcode `\^12\catcode `\_12\catcode `\%12\relax}%
\providecommand \@@startlink[1]{}%
\providecommand \@@endlink[0]{}%
\providecommand \url  [0]{\begingroup\@sanitize@url \@url }%
\providecommand \@url [1]{\endgroup\@href {#1}{\urlprefix }}%
\providecommand \urlprefix  [0]{URL }%
\providecommand \Eprint [0]{\href }%
\providecommand \doibase [0]{https://doi.org/}%
\providecommand \selectlanguage [0]{\@gobble}%
\providecommand \bibinfo  [0]{\@secondoftwo}%
\providecommand \bibfield  [0]{\@secondoftwo}%
\providecommand \translation [1]{[#1]}%
\providecommand \BibitemOpen [0]{}%
\providecommand \bibitemStop [0]{}%
\providecommand \bibitemNoStop [0]{.\EOS\space}%
\providecommand \EOS [0]{\spacefactor3000\relax}%
\providecommand \BibitemShut  [1]{\csname bibitem#1\endcsname}%
\let\auto@bib@innerbib\@empty
\bibitem [{\citenamefont {Bianconi}(2015)}]{Bianconi:2015}%
  \BibitemOpen
  \bibfield  {author} {\bibinfo {author} {\bibfnamefont {G.}~\bibnamefont
  {Bianconi}},\ }\href {https://doi.org/10.1209/0295-5075/111/56001} {\bibfield
   {journal} {\bibinfo  {journal} {{EPL} (Europhysics Letters)}\ }\textbf
  {\bibinfo {volume} {111}},\ \bibinfo {pages} {56001} (\bibinfo {year}
  {2015})}\BibitemShut {NoStop}%
\bibitem [{\citenamefont {Mahler}\ and\ \citenamefont
  {Weberru\ss{}}(1998)}]{Mahler:1998}%
  \BibitemOpen
  \bibfield  {author} {\bibinfo {author} {\bibfnamefont {G.}~\bibnamefont
  {Mahler}}\ and\ \bibinfo {author} {\bibfnamefont {V.~A.}\ \bibnamefont
  {Weberru\ss{}}},\ }\href {https://doi.org/10.1007/978-3-662-03669-3} {\emph
  {\bibinfo {title} {{Quantum Networks}}}}\ (\bibinfo  {publisher}
  {Springer-Verlag Berlin Heidelberg},\ \bibinfo {year} {1998})\BibitemShut
  {NoStop}%
\bibitem [{\citenamefont {Gisin}\ and\ \citenamefont
  {Thew}(2007)}]{Gisin:2007}%
  \BibitemOpen
  \bibfield  {author} {\bibinfo {author} {\bibfnamefont {N.}~\bibnamefont
  {Gisin}}\ and\ \bibinfo {author} {\bibfnamefont {R.}~\bibnamefont {Thew}},\
  }\href {https://doi.org/10.1038/nphoton.2007.22} {\bibfield  {journal}
  {\bibinfo  {journal} {Nature Photonics}\ }\textbf {\bibinfo {volume} {1}},\
  \bibinfo {pages} {165} (\bibinfo {year} {2007})}\BibitemShut {NoStop}%
\bibitem [{\citenamefont {Chen}(2021)}]{Chen:2021}%
  \BibitemOpen
  \bibfield  {author} {\bibinfo {author} {\bibfnamefont {J.}~\bibnamefont
  {Chen}},\ }\href {https://doi.org/10.1088/1742-6596/1865/2/022008} {\bibfield
   {journal} {\bibinfo  {journal} {Journal of Physics: Conference Series}\
  }\textbf {\bibinfo {volume} {1865}},\ \bibinfo {pages} {022008} (\bibinfo
  {year} {2021})}\BibitemShut {NoStop}%
\bibitem [{\citenamefont {Lambert}(2021)}]{Lambert2021}%
  \BibitemOpen
  \bibfield  {author} {\bibinfo {author} {\bibfnamefont {C.~J.}\ \bibnamefont
  {Lambert}},\ }\href {https://doi.org/10.1088/978-0-7503-3639-0} {\emph
  {\bibinfo {title} {{ Quantum Transport in Nanostructures and Molecules: An
  introduction to molecular electronics}}}}\ (\bibinfo  {publisher} {IOP
  Publishing},\ \bibinfo {year} {2021})\BibitemShut {NoStop}%
\bibitem [{\citenamefont {Beenakker}\ and\ \citenamefont {van
  {H}outen}(2008)}]{Beenakker:2008}%
  \BibitemOpen
  \bibfield  {author} {\bibinfo {author} {\bibfnamefont {C.~W.~J.}\
  \bibnamefont {Beenakker}}\ and\ \bibinfo {author} {\bibfnamefont
  {H.}~\bibnamefont {van {H}outen}},\ }\href
  {https://doi.org/https://doi.org/10.1016/S0081-1947(08)60091-0} {\bibfield
  {journal} {\bibinfo  {journal} {Solid State Physics}\ }\textbf {\bibinfo
  {volume} {44}},\ \bibinfo {pages} {1} (\bibinfo {year} {2008})}\BibitemShut
  {NoStop}%
\bibitem [{\citenamefont {Lambert}\ \emph {et~al.}(2013)\citenamefont
  {Lambert}, \citenamefont {Chen}, \citenamefont {Cheng}, \citenamefont {Li},
  \citenamefont {Chen},\ and\ \citenamefont {Nori}}]{Lambert_review:2013}%
  \BibitemOpen
  \bibfield  {author} {\bibinfo {author} {\bibfnamefont {N.}~\bibnamefont
  {Lambert}}, \bibinfo {author} {\bibfnamefont {Y.-N.}\ \bibnamefont {Chen}},
  \bibinfo {author} {\bibfnamefont {Y.-C.}\ \bibnamefont {Cheng}}, \bibinfo
  {author} {\bibfnamefont {C.-M.}\ \bibnamefont {Li}}, \bibinfo {author}
  {\bibfnamefont {G.-Y.}\ \bibnamefont {Chen}},\ and\ \bibinfo {author}
  {\bibfnamefont {F.}~\bibnamefont {Nori}},\ }\href
  {https://doi.org/10.1038/nphys2474} {\bibfield  {journal} {\bibinfo
  {journal} {Nature Physics}\ }\textbf {\bibinfo {volume} {9}},\ \bibinfo
  {pages} {10} (\bibinfo {year} {2013})}\BibitemShut {NoStop}%
\bibitem [{\citenamefont {Huelga}\ and\ \citenamefont
  {Plenio}(2013)}]{Huelga_Plenio_quantum_bio:2013}%
  \BibitemOpen
  \bibfield  {author} {\bibinfo {author} {\bibfnamefont {S.}~\bibnamefont
  {Huelga}}\ and\ \bibinfo {author} {\bibfnamefont {M.}~\bibnamefont
  {Plenio}},\ }\href {https://doi.org/10.1080/00405000.2013.829687} {\bibfield
  {journal} {\bibinfo  {journal} {Contemporary Physics}\ }\textbf {\bibinfo
  {volume} {54}},\ \bibinfo {pages} {181} (\bibinfo {year} {2013})}\BibitemShut
  {NoStop}%
\bibitem [{\citenamefont {Horodecki}\ \emph {et~al.}(2009)\citenamefont
  {Horodecki}, \citenamefont {Horodecki}, \citenamefont {Horodecki},\ and\
  \citenamefont {Horodecki}}]{Horodecki_rev:2009}%
  \BibitemOpen
  \bibfield  {author} {\bibinfo {author} {\bibfnamefont {R.}~\bibnamefont
  {Horodecki}}, \bibinfo {author} {\bibfnamefont {P.}~\bibnamefont
  {Horodecki}}, \bibinfo {author} {\bibfnamefont {M.}~\bibnamefont
  {Horodecki}},\ and\ \bibinfo {author} {\bibfnamefont {K.}~\bibnamefont
  {Horodecki}},\ }\href {https://doi.org/10.1103/RevModPhys.81.865} {\bibfield
  {journal} {\bibinfo  {journal} {Rev. Mod. Phys.}\ }\textbf {\bibinfo {volume}
  {81}},\ \bibinfo {pages} {865} (\bibinfo {year} {2009})}\BibitemShut
  {NoStop}%
\bibitem [{\citenamefont {Zurek}(2003)}]{Zurek_Rev:2003}%
  \BibitemOpen
  \bibfield  {author} {\bibinfo {author} {\bibfnamefont {W.~H.}\ \bibnamefont
  {Zurek}},\ }\href {https://doi.org/10.1103/RevModPhys.75.715} {\bibfield
  {journal} {\bibinfo  {journal} {Rev. Mod. Phys.}\ }\textbf {\bibinfo {volume}
  {75}},\ \bibinfo {pages} {715} (\bibinfo {year} {2003})}\BibitemShut
  {NoStop}%
\bibitem [{\citenamefont {Breuer}\ and\ \citenamefont
  {Petruccione}(2002)}]{Breuer-Petruccione}%
  \BibitemOpen
  \bibfield  {author} {\bibinfo {author} {\bibfnamefont {H.-P.}\ \bibnamefont
  {Breuer}}\ and\ \bibinfo {author} {\bibfnamefont {F.}~\bibnamefont
  {Petruccione}},\ }\href
  {https://doi.org/10.1093/acprof:oso/9780199213900.001.0001} {\emph {\bibinfo
  {title} {The Theory of Open Quantum Systems}}}\ (\bibinfo  {publisher}
  {Oxford University Press},\ \bibinfo {address} {Oxford},\ \bibinfo {year}
  {2002})\BibitemShut {NoStop}%
\bibitem [{\citenamefont {Rivas}\ and\ \citenamefont
  {Huelga}(2012)}]{Rivas2012}%
  \BibitemOpen
  \bibfield  {author} {\bibinfo {author} {\bibfnamefont {A.}~\bibnamefont
  {Rivas}}\ and\ \bibinfo {author} {\bibfnamefont {S.~F.}\ \bibnamefont
  {Huelga}},\ }\href {https://doi.org/10.1007/978-3-642-23354-8} {\emph
  {\bibinfo {title} {{Open Quantum Systems: An Introduction}}}},\
  SpringerBriefs in Physics\ (\bibinfo  {publisher} {Springer-Verlag Berlin
  Heidelberg},\ \bibinfo {year} {2012})\BibitemShut {NoStop}%
\bibitem [{\citenamefont {de~Vega}\ and\ \citenamefont
  {Alonso}(2017)}]{deVega_Alonso_rev:2017}%
  \BibitemOpen
  \bibfield  {author} {\bibinfo {author} {\bibfnamefont {I.}~\bibnamefont
  {de~Vega}}\ and\ \bibinfo {author} {\bibfnamefont {D.}~\bibnamefont
  {Alonso}},\ }\href {https://doi.org/10.1103/RevModPhys.89.015001} {\bibfield
  {journal} {\bibinfo  {journal} {Rev. Mod. Phys.}\ }\textbf {\bibinfo {volume}
  {89}},\ \bibinfo {pages} {015001} (\bibinfo {year} {2017})}\BibitemShut
  {NoStop}%
\bibitem [{\citenamefont {Engel}\ \emph {et~al.}(2007)\citenamefont {Engel},
  \citenamefont {Calhoun}, \citenamefont {Read}, \citenamefont {Ahn},
  \citenamefont {Man{\v c}al}, \citenamefont {Cheng}, \citenamefont
  {Blankenship},\ and\ \citenamefont {Fleming}}]{Engel:2007}%
  \BibitemOpen
  \bibfield  {author} {\bibinfo {author} {\bibfnamefont {G.~S.}\ \bibnamefont
  {Engel}}, \bibinfo {author} {\bibfnamefont {T.~R.}\ \bibnamefont {Calhoun}},
  \bibinfo {author} {\bibfnamefont {E.~L.}\ \bibnamefont {Read}}, \bibinfo
  {author} {\bibfnamefont {T.-K.}\ \bibnamefont {Ahn}}, \bibinfo {author}
  {\bibfnamefont {T.}~\bibnamefont {Man{\v c}al}}, \bibinfo {author}
  {\bibfnamefont {Y.-C.}\ \bibnamefont {Cheng}}, \bibinfo {author}
  {\bibfnamefont {R.~E.}\ \bibnamefont {Blankenship}},\ and\ \bibinfo {author}
  {\bibfnamefont {G.~R.}\ \bibnamefont {Fleming}},\ }\href
  {https://doi.org/10.1038/nature05678} {\bibfield  {journal} {\bibinfo
  {journal} {Nature}\ }\textbf {\bibinfo {volume} {446}},\ \bibinfo {pages}
  {782} (\bibinfo {year} {2007})}\BibitemShut {NoStop}%
\bibitem [{\citenamefont {Cheng}\ and\ \citenamefont
  {Fleming}(2009)}]{Fleming:2009}%
  \BibitemOpen
  \bibfield  {author} {\bibinfo {author} {\bibfnamefont {Y.-C.}\ \bibnamefont
  {Cheng}}\ and\ \bibinfo {author} {\bibfnamefont {G.~R.}\ \bibnamefont
  {Fleming}},\ }\href {https://doi.org/10.1146/annurev.physchem.040808.090259}
  {\bibfield  {journal} {\bibinfo  {journal} {Annual Review of Physical
  Chemistry}\ }\textbf {\bibinfo {volume} {60}},\ \bibinfo {pages} {241}
  (\bibinfo {year} {2009})},\ \bibinfo {note} {pMID: 18999996}\BibitemShut
  {NoStop}%
\bibitem [{\citenamefont {Jang}\ and\ \citenamefont
  {Mennucci}(2018)}]{LHC_rev:2018}%
  \BibitemOpen
  \bibfield  {author} {\bibinfo {author} {\bibfnamefont {S.~J.}\ \bibnamefont
  {Jang}}\ and\ \bibinfo {author} {\bibfnamefont {B.}~\bibnamefont
  {Mennucci}},\ }\href {https://doi.org/10.1103/RevModPhys.90.035003}
  {\bibfield  {journal} {\bibinfo  {journal} {Rev. Mod. Phys.}\ }\textbf
  {\bibinfo {volume} {90}},\ \bibinfo {pages} {035003} (\bibinfo {year}
  {2018})}\BibitemShut {NoStop}%
\bibitem [{\citenamefont {Plenio}\ and\ \citenamefont
  {Huelga}(2008)}]{Plenio:2008}%
  \BibitemOpen
  \bibfield  {author} {\bibinfo {author} {\bibfnamefont {M.~B.}\ \bibnamefont
  {Plenio}}\ and\ \bibinfo {author} {\bibfnamefont {S.~F.}\ \bibnamefont
  {Huelga}},\ }\href {https://doi.org/10.1088/1367-2630/10/11/113019}
  {\bibfield  {journal} {\bibinfo  {journal} {New Journal of Physics}\ }\textbf
  {\bibinfo {volume} {10}},\ \bibinfo {pages} {113019} (\bibinfo {year}
  {2008})}\BibitemShut {NoStop}%
\bibitem [{\citenamefont {Caruso}\ \emph {et~al.}(2009)\citenamefont {Caruso},
  \citenamefont {Chin}, \citenamefont {Datta}, \citenamefont {Huelga},\ and\
  \citenamefont {Plenio}}]{Caruso:2009}%
  \BibitemOpen
  \bibfield  {author} {\bibinfo {author} {\bibfnamefont {F.}~\bibnamefont
  {Caruso}}, \bibinfo {author} {\bibfnamefont {A.~W.}\ \bibnamefont {Chin}},
  \bibinfo {author} {\bibfnamefont {A.}~\bibnamefont {Datta}}, \bibinfo
  {author} {\bibfnamefont {S.~F.}\ \bibnamefont {Huelga}},\ and\ \bibinfo
  {author} {\bibfnamefont {M.~B.}\ \bibnamefont {Plenio}},\ }\href
  {https://doi.org/10.1063/1.3223548} {\bibfield  {journal} {\bibinfo
  {journal} {The Journal of Chemical Physics}\ }\textbf {\bibinfo {volume}
  {131}},\ \bibinfo {pages} {105106} (\bibinfo {year} {2009})}\BibitemShut
  {NoStop}%
\bibitem [{\citenamefont {Chin}\ \emph {et~al.}(2013)\citenamefont {Chin},
  \citenamefont {Prior}, \citenamefont {Rosenbach}, \citenamefont
  {Caycedo-Soler}, \citenamefont {Huelga},\ and\ \citenamefont
  {Plenio}}]{Chin:2013}%
  \BibitemOpen
  \bibfield  {author} {\bibinfo {author} {\bibfnamefont {A.~W.}\ \bibnamefont
  {Chin}}, \bibinfo {author} {\bibfnamefont {J.}~\bibnamefont {Prior}},
  \bibinfo {author} {\bibfnamefont {R.}~\bibnamefont {Rosenbach}}, \bibinfo
  {author} {\bibfnamefont {F.}~\bibnamefont {Caycedo-Soler}}, \bibinfo {author}
  {\bibfnamefont {S.~F.}\ \bibnamefont {Huelga}},\ and\ \bibinfo {author}
  {\bibfnamefont {M.~B.}\ \bibnamefont {Plenio}},\ }\href
  {https://doi.org/10.1038/nphys2515} {\bibfield  {journal} {\bibinfo
  {journal} {Nature Physics}\ }\textbf {\bibinfo {volume} {9}},\ \bibinfo
  {pages} {113} (\bibinfo {year} {2013})}\BibitemShut {NoStop}%
\bibitem [{\citenamefont {Anderson}(1958)}]{Anderson:1958}%
  \BibitemOpen
  \bibfield  {author} {\bibinfo {author} {\bibfnamefont {P.~W.}\ \bibnamefont
  {Anderson}},\ }\href {https://doi.org/10.1103/PhysRev.109.1492} {\bibfield
  {journal} {\bibinfo  {journal} {Phys. Rev.}\ }\textbf {\bibinfo {volume}
  {109}},\ \bibinfo {pages} {1492} (\bibinfo {year} {1958})}\BibitemShut
  {NoStop}%
\bibitem [{\citenamefont {Anderson}(1978)}]{Anderson_rev:1978}%
  \BibitemOpen
  \bibfield  {author} {\bibinfo {author} {\bibfnamefont {P.~W.}\ \bibnamefont
  {Anderson}},\ }\href {https://doi.org/10.1103/RevModPhys.50.191} {\bibfield
  {journal} {\bibinfo  {journal} {Rev. Mod. Phys.}\ }\textbf {\bibinfo {volume}
  {50}},\ \bibinfo {pages} {191} (\bibinfo {year} {1978})}\BibitemShut
  {NoStop}%
\bibitem [{\citenamefont {Mohseni}\ \emph {et~al.}(2008)\citenamefont
  {Mohseni}, \citenamefont {Rebentrost}, \citenamefont {Lloyd},\ and\
  \citenamefont {Aspuru-Guzik}}]{Mohseni:2008}%
  \BibitemOpen
  \bibfield  {author} {\bibinfo {author} {\bibfnamefont {M.}~\bibnamefont
  {Mohseni}}, \bibinfo {author} {\bibfnamefont {P.}~\bibnamefont {Rebentrost}},
  \bibinfo {author} {\bibfnamefont {S.}~\bibnamefont {Lloyd}},\ and\ \bibinfo
  {author} {\bibfnamefont {A.}~\bibnamefont {Aspuru-Guzik}},\ }\href
  {https://doi.org/10.1063/1.3002335} {\bibfield  {journal} {\bibinfo
  {journal} {The Journal of Chemical Physics}\ }\textbf {\bibinfo {volume}
  {129}},\ \bibinfo {pages} {174106} (\bibinfo {year} {2008})}\BibitemShut
  {NoStop}%
\bibitem [{\citenamefont {Rebentrost}\ \emph {et~al.}(2009)\citenamefont
  {Rebentrost}, \citenamefont {Mohseni}, \citenamefont {Kassal}, \citenamefont
  {Lloyd},\ and\ \citenamefont {Aspuru-Guzik}}]{Rebentrost:2009}%
  \BibitemOpen
  \bibfield  {author} {\bibinfo {author} {\bibfnamefont {P.}~\bibnamefont
  {Rebentrost}}, \bibinfo {author} {\bibfnamefont {M.}~\bibnamefont {Mohseni}},
  \bibinfo {author} {\bibfnamefont {I.}~\bibnamefont {Kassal}}, \bibinfo
  {author} {\bibfnamefont {S.}~\bibnamefont {Lloyd}},\ and\ \bibinfo {author}
  {\bibfnamefont {A.}~\bibnamefont {Aspuru-Guzik}},\ }\href
  {https://doi.org/10.1088/1367-2630/11/3/033003} {\bibfield  {journal}
  {\bibinfo  {journal} {New Journal of Physics}\ }\textbf {\bibinfo {volume}
  {11}},\ \bibinfo {pages} {033003} (\bibinfo {year} {2009})}\BibitemShut
  {NoStop}%
\bibitem [{\citenamefont {Adolphs}\ and\ \citenamefont
  {Renger}(2006)}]{Adolphs:2006}%
  \BibitemOpen
  \bibfield  {author} {\bibinfo {author} {\bibfnamefont {J.}~\bibnamefont
  {Adolphs}}\ and\ \bibinfo {author} {\bibfnamefont {T.}~\bibnamefont
  {Renger}},\ }\href
  {https://doi.org/https://doi.org/10.1529/biophysj.105.079483} {\bibfield
  {journal} {\bibinfo  {journal} {Biophysical Journal}\ }\textbf {\bibinfo
  {volume} {91}},\ \bibinfo {pages} {2778} (\bibinfo {year}
  {2006})}\BibitemShut {NoStop}%
\bibitem [{\citenamefont {Davidson}\ \emph {et~al.}(2021)\citenamefont
  {Davidson}, \citenamefont {Pollock},\ and\ \citenamefont
  {Gauger}}]{PhysRevResearch.3.L032001}%
  \BibitemOpen
  \bibfield  {author} {\bibinfo {author} {\bibfnamefont {S.}~\bibnamefont
  {Davidson}}, \bibinfo {author} {\bibfnamefont {F.~A.}\ \bibnamefont
  {Pollock}},\ and\ \bibinfo {author} {\bibfnamefont {E.}~\bibnamefont
  {Gauger}},\ }\href {https://doi.org/10.1103/PhysRevResearch.3.L032001}
  {\bibfield  {journal} {\bibinfo  {journal} {Phys. Rev. Research}\ }\textbf
  {\bibinfo {volume} {3}},\ \bibinfo {pages} {L032001} (\bibinfo {year}
  {2021})}\BibitemShut {NoStop}%
\bibitem [{\citenamefont {Albert}\ and\ \citenamefont
  {Barab\'asi}(2002)}]{Barabasi_rev:2002}%
  \BibitemOpen
  \bibfield  {author} {\bibinfo {author} {\bibfnamefont {R.}~\bibnamefont
  {Albert}}\ and\ \bibinfo {author} {\bibfnamefont {A.-L.}\ \bibnamefont
  {Barab\'asi}},\ }\href {https://doi.org/10.1103/RevModPhys.74.47} {\bibfield
  {journal} {\bibinfo  {journal} {Rev. Mod. Phys.}\ }\textbf {\bibinfo {volume}
  {74}},\ \bibinfo {pages} {47} (\bibinfo {year} {2002})}\BibitemShut {NoStop}%
\bibitem [{\citenamefont {Chin}\ \emph {et~al.}(2010)\citenamefont {Chin},
  \citenamefont {Rivas}, \citenamefont {Huelga},\ and\ \citenamefont
  {Plenio}}]{Chin:2010}%
  \BibitemOpen
  \bibfield  {author} {\bibinfo {author} {\bibfnamefont {A.~W.}\ \bibnamefont
  {Chin}}, \bibinfo {author} {\bibfnamefont {A.}~\bibnamefont {Rivas}},
  \bibinfo {author} {\bibfnamefont {S.~F.}\ \bibnamefont {Huelga}},\ and\
  \bibinfo {author} {\bibfnamefont {M.~B.}\ \bibnamefont {Plenio}},\ }\href
  {https://doi.org/10.1063/1.3490188} {\bibfield  {journal} {\bibinfo
  {journal} {Journal of Mathematical Physics}\ }\textbf {\bibinfo {volume}
  {51}},\ \bibinfo {pages} {092109} (\bibinfo {year} {2010})}\BibitemShut
  {NoStop}%
\bibitem [{\citenamefont {Prior}\ \emph {et~al.}(2010)\citenamefont {Prior},
  \citenamefont {Chin}, \citenamefont {Huelga},\ and\ \citenamefont
  {Plenio}}]{Prior:2010}%
  \BibitemOpen
  \bibfield  {author} {\bibinfo {author} {\bibfnamefont {J.}~\bibnamefont
  {Prior}}, \bibinfo {author} {\bibfnamefont {A.~W.}\ \bibnamefont {Chin}},
  \bibinfo {author} {\bibfnamefont {S.~F.}\ \bibnamefont {Huelga}},\ and\
  \bibinfo {author} {\bibfnamefont {M.~B.}\ \bibnamefont {Plenio}},\ }\href
  {https://doi.org/10.1103/PhysRevLett.105.050404} {\bibfield  {journal}
  {\bibinfo  {journal} {Phys. Rev. Lett.}\ }\textbf {\bibinfo {volume} {105}},\
  \bibinfo {pages} {050404} (\bibinfo {year} {2010})}\BibitemShut {NoStop}%
\bibitem [{\citenamefont {Tamascelli}\ \emph {et~al.}(2019)\citenamefont
  {Tamascelli}, \citenamefont {Smirne}, \citenamefont {Lim}, \citenamefont
  {Huelga},\ and\ \citenamefont {Plenio}}]{Tamascelli:2019}%
  \BibitemOpen
  \bibfield  {author} {\bibinfo {author} {\bibfnamefont {D.}~\bibnamefont
  {Tamascelli}}, \bibinfo {author} {\bibfnamefont {A.}~\bibnamefont {Smirne}},
  \bibinfo {author} {\bibfnamefont {J.}~\bibnamefont {Lim}}, \bibinfo {author}
  {\bibfnamefont {S.~F.}\ \bibnamefont {Huelga}},\ and\ \bibinfo {author}
  {\bibfnamefont {M.~B.}\ \bibnamefont {Plenio}},\ }\href
  {https://doi.org/10.1103/PhysRevLett.123.090402} {\bibfield  {journal}
  {\bibinfo  {journal} {Phys. Rev. Lett.}\ }\textbf {\bibinfo {volume} {123}},\
  \bibinfo {pages} {090402} (\bibinfo {year} {2019})}\BibitemShut {NoStop}%
\bibitem [{\citenamefont {Am-Shallem}\ \emph {et~al.}(2015)\citenamefont
  {Am-Shallem}, \citenamefont {Levy}, \citenamefont {Schaefer},\ and\
  \citenamefont {Kosloff}}]{vectorization:2015}%
  \BibitemOpen
  \bibfield  {author} {\bibinfo {author} {\bibfnamefont {M.}~\bibnamefont
  {Am-Shallem}}, \bibinfo {author} {\bibfnamefont {A.}~\bibnamefont {Levy}},
  \bibinfo {author} {\bibfnamefont {I.}~\bibnamefont {Schaefer}},\ and\
  \bibinfo {author} {\bibfnamefont {R.}~\bibnamefont {Kosloff}},\ }\href
  {https://doi.org/10.48550/ARXIV.1510.08634} {\bibinfo {title} {{Three
  approaches for representing Lindblad dynamics by a matrix-vector notation}}}
  (\bibinfo {year} {2015})\BibitemShut {NoStop}%
\bibitem [{\citenamefont {Abadi}\ \emph {et~al.}(2015)\citenamefont {Abadi},
  \citenamefont {Agarwal}, \citenamefont {Barham}, \citenamefont {Brevdo},
  \citenamefont {Chen}, \citenamefont {Citro}, \citenamefont {Corrado},
  \citenamefont {Davis}, \citenamefont {Dean}, \citenamefont {Devin},
  \citenamefont {Ghemawat}, \citenamefont {Goodfellow}, \citenamefont {Harp},
  \citenamefont {Irving}, \citenamefont {Isard}, \citenamefont {Jia},
  \citenamefont {Jozefowicz}, \citenamefont {Kaiser}, \citenamefont {Kudlur},
  \citenamefont {Levenberg}, \citenamefont {Man\'{e}}, \citenamefont {Monga},
  \citenamefont {Moore}, \citenamefont {Murray}, \citenamefont {Olah},
  \citenamefont {Schuster}, \citenamefont {Shlens}, \citenamefont {Steiner},
  \citenamefont {Sutskever}, \citenamefont {Talwar}, \citenamefont {Tucker},
  \citenamefont {Vanhoucke}, \citenamefont {Vasudevan}, \citenamefont
  {Vi\'{e}gas}, \citenamefont {Vinyals}, \citenamefont {Warden}, \citenamefont
  {Wattenberg}, \citenamefont {Wicke}, \citenamefont {Yu},\ and\ \citenamefont
  {Zheng}}]{tensorflow:2015}%
  \BibitemOpen
  \bibfield  {author} {\bibinfo {author} {\bibfnamefont {M.}~\bibnamefont
  {Abadi}}, \bibinfo {author} {\bibfnamefont {A.}~\bibnamefont {Agarwal}},
  \bibinfo {author} {\bibfnamefont {P.}~\bibnamefont {Barham}}, \bibinfo
  {author} {\bibfnamefont {E.}~\bibnamefont {Brevdo}}, \bibinfo {author}
  {\bibfnamefont {Z.}~\bibnamefont {Chen}}, \bibinfo {author} {\bibfnamefont
  {C.}~\bibnamefont {Citro}}, \bibinfo {author} {\bibfnamefont {G.~S.}\
  \bibnamefont {Corrado}}, \bibinfo {author} {\bibfnamefont {A.}~\bibnamefont
  {Davis}}, \bibinfo {author} {\bibfnamefont {J.}~\bibnamefont {Dean}},
  \bibinfo {author} {\bibfnamefont {M.}~\bibnamefont {Devin}}, \bibinfo
  {author} {\bibfnamefont {S.}~\bibnamefont {Ghemawat}}, \bibinfo {author}
  {\bibfnamefont {I.}~\bibnamefont {Goodfellow}}, \bibinfo {author}
  {\bibfnamefont {A.}~\bibnamefont {Harp}}, \bibinfo {author} {\bibfnamefont
  {G.}~\bibnamefont {Irving}}, \bibinfo {author} {\bibfnamefont
  {M.}~\bibnamefont {Isard}}, \bibinfo {author} {\bibfnamefont
  {Y.}~\bibnamefont {Jia}}, \bibinfo {author} {\bibfnamefont {R.}~\bibnamefont
  {Jozefowicz}}, \bibinfo {author} {\bibfnamefont {L.}~\bibnamefont {Kaiser}},
  \bibinfo {author} {\bibfnamefont {M.}~\bibnamefont {Kudlur}}, \bibinfo
  {author} {\bibfnamefont {J.}~\bibnamefont {Levenberg}}, \bibinfo {author}
  {\bibfnamefont {D.}~\bibnamefont {Man\'{e}}}, \bibinfo {author}
  {\bibfnamefont {R.}~\bibnamefont {Monga}}, \bibinfo {author} {\bibfnamefont
  {S.}~\bibnamefont {Moore}}, \bibinfo {author} {\bibfnamefont
  {D.}~\bibnamefont {Murray}}, \bibinfo {author} {\bibfnamefont
  {C.}~\bibnamefont {Olah}}, \bibinfo {author} {\bibfnamefont {M.}~\bibnamefont
  {Schuster}}, \bibinfo {author} {\bibfnamefont {J.}~\bibnamefont {Shlens}},
  \bibinfo {author} {\bibfnamefont {B.}~\bibnamefont {Steiner}}, \bibinfo
  {author} {\bibfnamefont {I.}~\bibnamefont {Sutskever}}, \bibinfo {author}
  {\bibfnamefont {K.}~\bibnamefont {Talwar}}, \bibinfo {author} {\bibfnamefont
  {P.}~\bibnamefont {Tucker}}, \bibinfo {author} {\bibfnamefont
  {V.}~\bibnamefont {Vanhoucke}}, \bibinfo {author} {\bibfnamefont
  {V.}~\bibnamefont {Vasudevan}}, \bibinfo {author} {\bibfnamefont
  {F.}~\bibnamefont {Vi\'{e}gas}}, \bibinfo {author} {\bibfnamefont
  {O.}~\bibnamefont {Vinyals}}, \bibinfo {author} {\bibfnamefont
  {P.}~\bibnamefont {Warden}}, \bibinfo {author} {\bibfnamefont
  {M.}~\bibnamefont {Wattenberg}}, \bibinfo {author} {\bibfnamefont
  {M.}~\bibnamefont {Wicke}}, \bibinfo {author} {\bibfnamefont
  {Y.}~\bibnamefont {Yu}},\ and\ \bibinfo {author} {\bibfnamefont
  {X.}~\bibnamefont {Zheng}},\ }\href {https://www.tensorflow.org/} {\bibinfo
  {title} {{TensorFlow}: Large-scale machine learning on heterogeneous
  systems}} (\bibinfo {year} {2015}),\ \bibinfo {note} {software available from
  tensorflow.org}\BibitemShut {NoStop}%
\bibitem [{\citenamefont
  {Ruder}(2016)}]{https://doi.org/10.48550/arxiv.1609.04747}%
  \BibitemOpen
  \bibfield  {author} {\bibinfo {author} {\bibfnamefont {S.}~\bibnamefont
  {Ruder}},\ }\href {https://doi.org/10.48550/ARXIV.1609.04747} {\bibinfo
  {title} {An overview of gradient descent optimization algorithms}} (\bibinfo
  {year} {2016})\BibitemShut {NoStop}%
\bibitem [{\citenamefont {Tronrud}\ \emph {et~al.}(2009)\citenamefont
  {Tronrud}, \citenamefont {Wen}, \citenamefont {Gay},\ and\ \citenamefont
  {Blankenship}}]{Tronrud:2009}%
  \BibitemOpen
  \bibfield  {author} {\bibinfo {author} {\bibfnamefont {D.~E.}\ \bibnamefont
  {Tronrud}}, \bibinfo {author} {\bibfnamefont {J.}~\bibnamefont {Wen}},
  \bibinfo {author} {\bibfnamefont {L.}~\bibnamefont {Gay}},\ and\ \bibinfo
  {author} {\bibfnamefont {R.~E.}\ \bibnamefont {Blankenship}},\ }\href
  {https://doi.org/10.1007/s11120-009-9430-6} {\bibfield  {journal} {\bibinfo
  {journal} {Photosynthesis Research}\ }\textbf {\bibinfo {volume} {100}},\
  \bibinfo {pages} {79} (\bibinfo {year} {2009})}\BibitemShut {NoStop}%
\bibitem [{\citenamefont {Schmidt~am Busch}\ \emph {et~al.}(2011)\citenamefont
  {Schmidt~am Busch}, \citenamefont {M{\"u}h}, \citenamefont
  {El-Amine~Madjet},\ and\ \citenamefont {Renger}}]{amBusch:2011}%
  \BibitemOpen
  \bibfield  {author} {\bibinfo {author} {\bibfnamefont {M.}~\bibnamefont
  {Schmidt~am Busch}}, \bibinfo {author} {\bibfnamefont {F.}~\bibnamefont
  {M{\"u}h}}, \bibinfo {author} {\bibfnamefont {M.}~\bibnamefont
  {El-Amine~Madjet}},\ and\ \bibinfo {author} {\bibfnamefont {T.}~\bibnamefont
  {Renger}},\ }\bibfield  {booktitle} {\emph {\bibinfo {booktitle} {The Journal
  of Physical Chemistry Letters}},\ }\href {https://doi.org/10.1021/jz101541b}
  {\bibfield  {journal} {\bibinfo  {journal} {The Journal of Physical Chemistry
  Letters}\ }\textbf {\bibinfo {volume} {2}},\ \bibinfo {pages} {93} (\bibinfo
  {year} {2011})}\BibitemShut {NoStop}%
\bibitem [{\citenamefont {Misra}\ and\ \citenamefont
  {Sudarshan}(1977)}]{Misra:1977}%
  \BibitemOpen
  \bibfield  {author} {\bibinfo {author} {\bibfnamefont {B.}~\bibnamefont
  {Misra}}\ and\ \bibinfo {author} {\bibfnamefont {E.~C.~G.}\ \bibnamefont
  {Sudarshan}},\ }\href {https://doi.org/10.1063/1.523304} {\bibfield
  {journal} {\bibinfo  {journal} {Journal of Mathematical Physics}\ }\textbf
  {\bibinfo {volume} {18}},\ \bibinfo {pages} {756} (\bibinfo {year}
  {1977})}\BibitemShut {NoStop}%
\bibitem [{\citenamefont {Peres}(1980)}]{Peres:1980}%
  \BibitemOpen
  \bibfield  {author} {\bibinfo {author} {\bibfnamefont {A.}~\bibnamefont
  {Peres}},\ }\href {https://doi.org/10.1119/1.12204} {\bibfield  {journal}
  {\bibinfo  {journal} {American Journal of Physics}\ }\textbf {\bibinfo
  {volume} {48}},\ \bibinfo {pages} {931} (\bibinfo {year} {1980})}\BibitemShut
  {NoStop}%
\bibitem [{\citenamefont {Facchi}\ and\ \citenamefont
  {Pascazio}(2002)}]{Facchi:2002}%
  \BibitemOpen
  \bibfield  {author} {\bibinfo {author} {\bibfnamefont {P.}~\bibnamefont
  {Facchi}}\ and\ \bibinfo {author} {\bibfnamefont {S.}~\bibnamefont
  {Pascazio}},\ }\href {https://doi.org/10.1103/PhysRevLett.89.080401}
  {\bibfield  {journal} {\bibinfo  {journal} {Phys. Rev. Lett.}\ }\textbf
  {\bibinfo {volume} {89}},\ \bibinfo {pages} {080401} (\bibinfo {year}
  {2002})}\BibitemShut {NoStop}%
\bibitem [{\citenamefont {Baumgratz}\ \emph {et~al.}(2014)\citenamefont
  {Baumgratz}, \citenamefont {Cramer},\ and\ \citenamefont
  {Plenio}}]{Baumgratz:2014}%
  \BibitemOpen
  \bibfield  {author} {\bibinfo {author} {\bibfnamefont {T.}~\bibnamefont
  {Baumgratz}}, \bibinfo {author} {\bibfnamefont {M.}~\bibnamefont {Cramer}},\
  and\ \bibinfo {author} {\bibfnamefont {M.~B.}\ \bibnamefont {Plenio}},\
  }\href {https://doi.org/10.1103/PhysRevLett.113.140401} {\bibfield  {journal}
  {\bibinfo  {journal} {Phys. Rev. Lett.}\ }\textbf {\bibinfo {volume} {113}},\
  \bibinfo {pages} {140401} (\bibinfo {year} {2014})}\BibitemShut {NoStop}%
\end{thebibliography}%

\appendix
\section{Optimal Hamiltonians}\label{appendix0}
In this Appendix, we report some of the optimal Hamiltonians found during our optimizations and analysis. All energy values are expressed in units  of $1.2414 \times 10^{-4} \ \text{eV}$.

The optimal Hamiltonians, expressed in matrix form, found with the interactions described by \Cref{eq:interactions} for the results presented in Table \ref{tab:0vsOpt} in are

\begin{equation}
\begin{split}
\begin{pmatrix}
65.7 & -104.1 & 5.1 & -4.3 & 4.7 & -15.1 & -7.8\\
-104.1 & -11.1 & 32.6 & 7.1 & 5.4 & 8.3 & 0.8\\
5.1 & 32.6 & -56.1 & -46.8 & 1.0 & -8.1 & 5.1\\
-4.3 & 7.1 & -46.8 & -36.2 & -70.7 & -14.7 & -61.5\\
4.7 & 5.4 & 1.0 & -70.7 & -30.6 & 89.7 & -2.5\\
-15.1 & 8.3 & -8.1 & -14.7 & 89.7 & 55.7 & 32.7\\
-7.8 & 0.8 & 5.1 & -61.5 & -2.5 & 32.7 & 4.2
\end{pmatrix},
\end{split}
\end{equation}

for $\vec{\gamma}=\vec{\gamma}_{\text{ref}}$,

\begin{equation}
\begin{split}
\begin{pmatrix}
43.5 & -104.1 & 5.1 & -4.3 & 4.7 & -15.1 & -7.8\\
-104.1 & 13.7 & 32.6 & 7.1 & 5.4 & 8.3 & 0.8\\
5.1 & 32.6 & -45.8 & -46.8 & 1.0 & -8.1 & 5.1\\
-4.3 & 7.1 & -46.8 & -4.3 & -70.7 & -14.7 & -61.5\\
4.7 & 5.4 & 1.0 & -70.7 & -19.5 & 89.7 & -2.5\\
-15.1 & 8.3 & -8.1 & -14.7 & 89.7 & 14.4 & 32.7\\
-7.8 & 0.8 & 5.1 & -61.5 & -2.5 & 32.7 & -8.6
\end{pmatrix},
\end{split}
\end{equation}

for $\vec{\gamma}=\vec{1}$, and

\begin{equation}
\begin{split}
\begin{pmatrix}
-13.2 & -104.1 & 5.1 & -4.3 & 4.7 & -15.1 & -7.8\\
-104.1 & -1.7 & 32.6 & 7.1 & 5.4 & 8.3 & 0.8\\
5.1 & 32.6 & 16.1 & -46.8 & 1.0 & -8.1 & 5.1\\
-4.3 & 7.1 & -46.8 & -43.3 & -70.7 & -14.7 & -61.5\\
4.7 & 5.4 & 1.0 & -70.7 & 424.2 & 89.7 & -2.5\\
-15.1 & 8.3 & -8.1 & -14.7 & 89.7 & -568.8 & 32.7\\
-7.8 & 0.8 & 5.1 & -61.5 & -2.5 & 32.7 & 39.5
\end{pmatrix},
\end{split}
\end{equation}

for $\vec{\gamma}=\vec{0}$. 

While a simple intuition of the pattern of optimal matrix entries found in such examples seems to be elusive, one can notice that, in the absence of dephasing, the moduli of two of the optimized local energies are significantly higher than the rest. In the analysis reported in Section \ref{resilience} we found fully connected networks that achieved near perfect transfer (see Table \ref{tab:different_connections}) for different $\vec{\gamma}$. The corresponding Hamiltonians are

\begin{equation}
\begin{split}
\begin{pmatrix}
65.7 & 182.4 & -83.3 & -106.2 & -191.6 & -18.8 & -20.8 \\
182.4 & -11.1 & -152.6 & 91.9 & -162.3 & 55.6 & 183.7 \\
-83.3 & -152.6 & -56.1 & -132.0 & -190.2 & 177.3 & -101.7 \\
-106.2 & 91.9 & -132.0 & -36.2 & -161.3 & -169.0 & 144.6 \\
-191.6 & -162.3 & -190.2 & -161.3 & -30.6 & -106.3 & -102.8 \\
-18.8 & 55.6 & 177.3 & -169.0 & -106.3 & 55.7 & -111.5 \\
-20.8 & 183.7 & -101.7 & 144.6 & -102.8 & -111.5 & 4.2
\end{pmatrix},
\end{split}
\end{equation}

for $\vec{\gamma}=\vec{\gamma}_{\text{ref}}$,

\begin{equation}
\begin{split}
\begin{pmatrix}
43.5 & 102.9 & 92.8 & 63.8 & 28.7 & -136.6 & -183.1 \\
102.9 & 13.7 & 0.8 & 75.5 & -118.5 & 177.5 & 110.7 \\
92.8 & 0.8 & -45.8 & -140.9 & -198.8 & 134.9 & 144.7 \\
63.8 & 75.5 & -140.9 & -4.3 & -184.6 & -14.5 & -139.5 \\
28.7 & -118.5 & -198.8 & -184.6 & -19.5 & -153.8 & 5.2 \\
-136.6 & 177.5 & 134.9 & -14.5 & -153.8 & 14.4 & -188.0 \\
-183.1 & 110.7 & 144.7 & -139.5 & 5.2 & -188.0 & -8.6
\end{pmatrix},
\end{split}
\end{equation}

for $\vec{\gamma}=\vec{1}$, and

\begin{equation}
\begin{split}
\begin{pmatrix}
-13.2 & 6.5 & 64.0 & -147.1 & -71.3 & -46.6 & -156.7 \\
6.5 & -1.7 & 13.2 & 20.5 & -102.7 & 68.0 & 56.6 \\
64.0 & 13.2 & 16.1 & -164.0 & 154.5 & 95.7 & -187.0 \\
-147.1 & 20.5 & -164.0 & -43.3 & -86.8 & -17.3 & 43.4 \\
-71.3 & -102.7 & 154.5 & -86.8 & 424.2 & 72.0 & 70.9 \\
-46.6 & 68.0 & 95.7 & -17.3 & 72.0 & -568.8 & 155.8 \\
-156.7 & 56.6 & -187.0 & 43.4 & 70.9 & 155.8 & 39.5
\end{pmatrix},
\end{split}
\end{equation}

for $\vec{\gamma}=\vec{0}$. Again, no  intuition for the optimality of such configurations is apparte. However, that many of the interactions are significantly stronger than in the FMO complex model described by \Cref{eq:interactions}.

\end{document}